\documentclass[a4paper,fleqn]{cas-sc}

\usepackage[authoryear]{natbib}
\usepackage{amsmath,amssymb}
\usepackage{amsthm}
\usepackage{graphicx}
\usepackage{booktabs}
\usepackage{multirow}
\usepackage{array}
\usepackage{longtable}
\usepackage{tabularx}
\usepackage{xcolor}
\usepackage{url}
\usepackage{hyperref}
\hypersetup{
    colorlinks=true,
    linkcolor=black,
    citecolor=blue!60!black,
    urlcolor=blue!60!black,
}
\usepackage{cleveref}
\usepackage{enumitem}
\usepackage{float}
\usepackage{placeins}
\usepackage{caption}
\usepackage{subcaption}
\usepackage{siunitx}
\usepackage{threeparttable}
\usepackage{pdflscape}

\begin{document}
\let\WriteBookmarks\relax
\def\floatpagepagefraction{1}
\def\textpagefraction{.001}

\shorttitle{Semantic structure and phantom collaborators in transportation research}
\shortauthors{S. Choi}

\title[mode=title]{Beyond coauthorship: semantic structure and phantom
  collaborators in transportation research, 1967--2025}

\author[1]{Seongjin Choi}[orcid=0000-0001-7140-537X]
\cormark[1]
\ead{chois@umn.edu}
\credit{Conceptualization, Methodology, Software, Data curation,
  Formal analysis, Writing -- original draft, Writing -- review and
  editing, Visualization, Funding acquisition}

\address[1]{Department of Civil, Environmental, and Geo-Engineering,
  University of Minnesota Twin Cities, 500 Pillsbury Dr. SE,
  Minneapolis, MN 55455, USA}

\cortext[cor1]{Corresponding author.}


\begin{abstract}
We present a semantic-structural atlas of transportation research built from
120{,}323 papers across 34 peer-reviewed journals published between 1967 and
2025, roughly an order of magnitude larger than and a decade beyond Sun and
Rahwan's~(2017) coauthorship study. We use OpenAlex and Crossref as open,
CC0-licensed data sources, resolve author identity through OpenAlex author
IDs, ORCID records, and manual alias resolution, and embed every paper with
SPECTER2 with Arora-style whitening concatenated with concept TF--IDF and
venue linear-discriminant projections. On this substrate we report three
findings. First, Leiden on the author-level semantic k-nearest-neighbor
graph yields 23 topic communities that agree only weakly with the 172
coauthor communities (normalized mutual information $0.23$), opening room
for a predictive layer that neither source encodes alone. Second, a
multiplex Leiden partition combining both edge types recovers 181
communities and localizes where collaboration and topic structure decouple. Third---the paper's core methodological contribution---we define \emph{phantom collaborators}, pairs of authors who are top-$K$ semantic neighbors yet $\geq 3$ hops apart in the coauthor graph, and show via a temporal hold-out (training cutoff 2019) that phantom pairs become real coauthors in 2020--2025 at a rate $16$ to $33$ times above random, popularity-weighted, and same-venue baselines, with a $68$-fold monotone gradient between the highest- and lowest-similarity buckets. All artifacts are released as a live, reproducible web atlas at \href{https://choi-seongjin.github.io/transport-atlas/}{choi-seongjin.github.io/transport-atlas}.
\end{abstract}


\maketitle


\section{Introduction}
\label{sec:intro}

Transportation research has grown into a large and tightly collaborative
research community. The reference snapshot is
\citet{sun2017coauthorship}, who analyzed roughly eleven thousand papers
across twenty transportation journals between $1990$ and $2015$ and
documented the community structure, centrality, and productivity
distribution of the coauthor network at that scale. A decade later the
field is roughly an order of magnitude larger. Major new venues have
emerged, multi-institution collaborations dominate
the top citation tail, and researchers trained in adjacent disciplines
such as ML, energy systems, and public health have entered the field
through automated-driving, energy-systems, and health-equity
submissions. A
refreshed bibliometric picture is therefore needed, both to calibrate
policy instruments such as funding panels and editorial reviews and to
give early-career researchers a navigable map of where the field has
arrived.

Most existing bibliometric studies in transportation rely on coauthor
and co-citation networks alone
\citep{sun2017coauthorship, jiang2020trb, modak2019fifty}. These
approaches capture who collaborates with whom, which is informative but
incomplete. Coauthorship structure tells us who has already worked
together. It does not reveal who \emph{would} work together on the same
problem if they met. However, paper embedding models now make that
second question tractable. Citation-aware pretrained language models
such as SPECTER \citep{cohan2020specter} and SPECTER2
\citep{singh2023specter2} project a paper's title and abstract into a
dense vector whose pairwise cosine similarity approximates topical
proximity. Aggregating these
vectors per author yields a semantic representation of a researcher's
oeuvre that is largely independent of the social ties in the coauthor
graph.

However, combining these two signals is not automatic. A paper-embedding
model trained on a single domain exhibits strong anisotropy that
compresses pairwise cosines to a narrow band
\citep{arora2017simple}. Community detection must therefore be designed
around the geometry of the embedding, not against it. Furthermore, the
predictive value of semantic proximity as a signal for \emph{future}
coauthorship has not been empirically tested on a transportation corpus
at scale. The
gap we address is the joint one. We lack a unified substrate that (i)
matches \citet{sun2017coauthorship} in descriptive rigour at ten times
the scale, (ii) adds a disciplined semantic layer on top of the
coauthor graph, and (iii) supports a predictive test of whether
semantic proximity carries collaboration-relevant signal beyond what
the coauthor graph already encodes.

To address these challenges, we build the \textbf{Transport Atlas}, a
static, browsable atlas of $\num{120323}$ papers across $34$ venues from
$1967$ to $2025$, and we organize the supporting analyses around three
research questions.

\begin{description}
  \item[\textbf{RQ1.}] How has the scale and topology of transportation
        coauthorship changed since \citet{sun2017coauthorship}?
  \item[\textbf{RQ2.}] Does topical structure, as recovered from
        paper-embedding proximity, agree with the social structure
        recovered from coauthorship alone?
  \item[\textbf{RQ3.}] Does semantic proximity predict future
        coauthorship in a temporally held-out evaluation, and if so
        does the signal remain after controlling for popularity and
        for shared publication venue?
\end{description}

\paragraph{Contributions.} The key contributions of this paper are as
follows, ordered from descriptive scaffolding to predictive payoff.

\begin{itemize}
  \item \textbf{We assemble a refreshed transportation corpus and
        extend \citet{sun2017coauthorship}'s descriptive scaffold to
        it.} The open corpus contains $\num{120323}$ papers across
        $34$ venues from $1967$ to $2025$ --- roughly an order of
        magnitude larger than the original --- with dedupe on DOI plus
        fuzzy title matching, author resolution through OpenAlex and
        ORCID, and citation counts reconciled as the max of OpenAlex
        and Crossref counts. We replicate their findings where
        comparable and report where the field has diverged.
  \item \textbf{We pair the coauthor network with a semantic network
        of researchers.} Using SPECTER2 paper embeddings aggregated to
        the author level, we build an author-level semantic graph that
        captures \emph{who writes about similar problems}, alongside
        the classical coauthor graph that captures \emph{who has
        already worked together}. The two graphs share authors as
        nodes but encode different relations, and either is unavailable
        in coauthorship-only studies.
  \item \textbf{We compare the two networks and find that they
        decouple.} Leiden on the semantic graph recovers $23$ topic
        communities; Leiden on the coauthor graph recovers $172$.
        Their normalized mutual information is only $0.23$. A multiplex Leiden
        partition combining both layers recovers $181$ joint
        communities and localizes where the two disagree.
  \item \textbf{We use the gap between the two networks to introduce
        \emph{phantom collaborators}.} Phantom collaborators --- pairs
        of authors who are top-$K$ semantic neighbors yet $\geq 3$
        hops apart in the coauthor graph --- become future coauthors
        $16$ to $33$ times more often than random,
        popularity-weighted, and same-venue baselines.
  \item \textbf{We release everything.} The atlas is live at
        \url{https://choi-seongjin.github.io/transport-atlas/} and
        the analysis pipeline at
        \url{https://github.com/UMN-Choi-Lab/transport-atlas}.
\end{itemize}

The rest of the paper is structured as follows.
Section~\ref{sec:related} reviews related work on transportation
bibliometrics, paper embeddings, and community detection.
Section~\ref{sec:data} describes the corpus and the dedupe,
disambiguation, and citation-reconciliation pipeline.
Section~\ref{sec:descriptive} presents the descriptive bibliometrics 
Section~\ref{sec:coauthor} analyzes the coauthor network structure, and
Sections~\ref{sec:semantic} and \ref{sec:multiplex} add the semantic and
multiplex layers.
Section~\ref{sec:phantom} develops the phantom-collaborator predictive
test, which is the paper's core methodological contribution.
Section~\ref{sec:trajectories} introduces a complementary career-arc
taxonomy that classifies each author's centroid trajectory by shape
(stayer, drifter, returner, switcher) and grounds the four classes
in named exemplars. Section~\ref{sec:atlas} showcases the atlas tool,
and Sections~\ref{sec:discussion}--\ref{sec:conclusion} discuss
limitations and conclude.

\section{Related work}
\label{sec:related}


\paragraph{Transportation bibliometrics.}
The most directly comparable work is \citet{sun2017coauthorship}, who built
a coauthorship network from about 11{,}000 papers across 20 transportation
journals between 1990 and 2015. They computed centrality, Louvain
communities, and a regression of citation impact on network position,
finding that productive authors benefit from embedding in geographically
dispersed coauthor networks and from publishing with highly cited
collaborators. \citet{jiang2020trb} and \citet{modak2019fifty} extend this
line with venue-specific analyses. \citet{jiang2020trb} covers $2{,}697$
TR-B papers from $1979$ to $2019$, and \citet{modak2019fifty} covers the
$10{,}449$-paper, $1967$--$2016$ seven-journal TR series, both using
VOSviewer \citep{vaneck2010vosviewer}. \Cref{tab:related-comparison}
summarizes these prior studies alongside our corpus. We build on them by
scaling the corpus roughly tenfold, expanding beyond TR-family journals
to the IEEE intelligent-transportation venues and TRR, using open data
sources in place of Web of Science, and adding a
semantic-embedding layer that prior coauthorship-only studies cannot
express.

\begin{table}[width=0.99\textwidth, pos = t]
  \centering
  \small
  \caption{Comparison of this paper with prior bibliometric studies of
    the transportation literature.}
  \label{tab:related-comparison}
    \begin{tabular}{llp{2cm}lp{3.2cm}p{3cm}}
  \toprule
  \textbf{Study} & \textbf{Papers} & \textbf{Venues} & \textbf{Window} & \textbf{Method} & \textbf{Data source} \\
  \midrule
  \citet{sun2017coauthorship} & 11{,}000 & 20 & 1990--2015 & Coauthor network, Louvain, citation regression & Web of Science \\
  \citet{jiang2020trb} & 2{,}697 & 1 (TR-B) & 1979--2019 & VOSviewer bibliometric & Web of Science \\
  \citet{modak2019fifty} & 10{,}449 & 7 (TR series) & 1967--2016 & VOSviewer bibliometric & Web of Science \\
  \midrule
  \textbf{This paper} & \textbf{\num{120323}} & \textbf{34} & \textbf{1967--2025} & \textbf{Coauthor + semantic + multiplex Leiden + phantom-collaborator prediction + career arc taxonomy} & \textbf{OpenAlex + Crossref + IEEE Xplore + Elsevier} \\
  \bottomrule
  \end{tabular}
\end{table}

\paragraph{Paper embeddings and community detection.}
Contemporary paper-embedding models
\citep{cohan2020specter,singh2023specter2} project scholarly documents into
dense vector spaces. In mono-domain corpora such as transportation research,
a strong shared ``domain direction'' dominates the pairwise-cosine
distribution. We remove it via the all-but-the-top transformation of
\citet{arora2017simple}. For community detection we prefer Leiden
\citep{traag2019leiden} over Louvain on the grounds of guaranteed
well-connectedness, and we combine coauthor and semantic-$k$NN edges into a
multiplex partition following \citet{mucha2010community}.

\paragraph{Link prediction framing.}
Section~\ref{sec:phantom} operationalises \emph{phantom collaborators} as a
temporal link-prediction problem in the sense of
\citet{libennowell2007link}. Unlike classical link-prediction features
(common neighbors, Adamic--Adar, Katz), our predictor is the cosine
similarity of author-level SPECTER2 centroids, which is largely
independent of network topology. That independence is precisely what
allows the prediction to surface \emph{latent} rather than
locally-obvious collaborations.

\section{Corpus and representation}
\label{sec:data}

\subsection{Corpus}
\label{subsec:data-corpus}

The atlas indexes \num{120323} papers across 34 peer-reviewed venues
published between 1967 and 2025. \Cref{tab:corpus-summary} reports headline
quantities. \Cref{tab:app-venues} (Appendix~\ref{app:venues}) lists each
venue's ISSN, paper count, and observed year range.

\begin{table}[width=0.99\textwidth, pos = t]
  \centering
  \caption{Corpus summary.}
  \label{tab:corpus-summary}
    \begin{tabular}{lr}
  \toprule
  \textbf{Quantity} & \textbf{Value} \\
  \midrule
  Papers & 120,323 \\
  Venues & 34 \\
  Year range & 1967--2025 \\
  Unique authors (deduped) & 168,090 \\
  Authors with ORCID & 56.9\% \\
  Total citations & 4,153,038 \\
  \bottomrule
  \end{tabular}
\end{table}

\subsection{Venue selection}
\label{subsec:data-venues}

Our goal in selecting venues is to cover the peer-reviewed journal
literature that a working transportation researcher would recognize as
core, while remaining ingestible from an open, redistributable source.
We start from the Clarivate Journal Citation Reports
\emph{Transportation} and \emph{Transportation Science \& Technology}
categories, which together cover $62$ and $77$ journals respectively in
the $2024$ edition. \citet{sun2017coauthorship} used the
\emph{Transportation} category alone, whereas \citet{jiang2020trb} and
\citet{modak2019fifty} focused on the TR family.

We impose four inclusion criteria. First, the venue must be a
peer-reviewed journal, magazine, or open-journal publication indexed in
Scopus, Web of Science, or both. Conference proceedings such as the IEEE
Intelligent Vehicles Symposium and the IEEE Intelligent Transportation
Systems Conference are therefore excluded. Second, the venue must be
covered by OpenAlex with at least $80\%$ DOI completeness, so our
downstream dedupe and citation-merging pipelines work uniformly. Third,
the scope must be transportation-research in the sense of the TR-family
editorial scopes, the IEEE intelligent-transportation venues, or their
direct methodological neighbors. This criterion excludes pure
aviation-operations journals (e.g.\ Journal of Air Transport
Management), pure maritime journals (Maritime Economics \& Logistics,
Maritime Transport Research), and narrow engineering journals
(SAE subseries, Proceedings of the Institution of Mechanical Engineers
subseries) because their coauthor and semantic structure is largely
disjoint from the rest of the corpus. Fourth, we cover the vehicle-technology and electrification side of
transportation research, which is underrepresented in
\citet{sun2017coauthorship} because the category membership of its
constituent journals tilted engineering rather than
transportation-research in the $1990$s and $2000$s. The seven venues
added here span a wide range of founding dates rather than a single
emergence window: Computer-Aided Civil and Infrastructure Engineering
(founded $1986$), IEEE Transactions on Vehicular Technology (indexed in
its current form from the $1950$s), IEEE Vehicular Technology Magazine
($2006$), Vehicular Communications ($2014$), IEEE Transactions on
Transportation Electrification ($2015$), eTransportation ($2020$), and
Green Energy and Intelligent Transportation ($2022$). What they share is
a sharp post-$2015$ growth in connected, automated, and electrified
vehicle research. The resulting $34$-venue list, enumerated in
Appendix~\ref{app:venues}.

\subsection{Ingest and deduplication}
\label{subsec:data-ingest}

We treat OpenAlex \citep{priem2022openalex} as the primary source of
bibliographic metadata for every venue in the corpus. Its CC0 license
permits redistribution of derived datasets, its coverage of our target
ISSNs is effectively complete back to 1967, and its records carry both
ORCID identifiers and stable OpenAlex author IDs that anchor the
disambiguation pipeline described in Section~\ref{subsec:data-authors}.
We resolve each venue by ISSN to an OpenAlex source ID and then paginate
works via cursor, checkpointing per-venue so that interrupted runs
resume without re-downloading.

Two sources enrich the primary feed. IEEE Xplore supplies abstracts and
author keywords for the IEEE Transactions on Intelligent Transportation
Systems, Transactions on Intelligent Vehicles, Intelligent Transportation
Systems Magazine, and the Open Journal of Intelligent Transportation
Systems, where OpenAlex abstract coverage is sparse before roughly 2010.
Crossref \citep{hendricks2020crossref} serves as a DOI fallback
for the small fraction of papers absent from OpenAlex.

Deduplication runs in two passes. We first collapse records sharing a
normalized DOI, which resolves the overwhelming majority of duplicates
introduced by multi-source ingest. Papers lacking a DOI or carrying
differing DOIs across sources pass through a fuzzy stage that requires
rapidfuzz token-set ratio above 95 on the normalized title, identical
publication year, and at least one overlapping author surname. We use
token-set rather than token-sort so that trailing fragments such as
``(Extended Abstract)'' or ``Part II'' do not defeat an otherwise valid
match. The conjunction of year and surname overlap guards against
the false positives that fuzzy-title matching alone would produce on
common transportation terminology.

\subsection{Author disambiguation}
\label{subsec:data-authors}

Author disambiguation proceeds through a three-level fallback chain. We
prefer the OpenAlex author ID when present, which covers essentially
every author-paper pair in the corpus and already reflects substantial
upstream cleaning. Where the OpenAlex ID is missing or where two IDs
plainly refer to the same researcher, we fall back to the ORCID
identifier, which is populated for roughly $55.7\%$ of author-paper pairs
(corresponding to $56.9\%$ of unique authors after deduplication, the
figure reported in Table~\ref{tab:corpus-summary}), and whose share has
risen steadily since the mid-2010s. As a final
fallback we use a canonical normalized name composed of the surname and
the first initial, with accents folded and punctuation stripped.

On top of this chain, an automatic ORCID-based merger consolidates
OpenAlex author IDs that resolve to the same ORCID under different
canonical names, producing several thousand merges that fold stub
records for ORCID-registered researchers into their primary identity.
A five-way split of \textit{Sun, Lijun} across OpenAlex IDs, caught
during development, has since been folded into the resolution
pipeline and now serves as a regression test.

A caveat on treating ORCID as ground truth: OpenAlex's ORCID-to-author
assignment is itself occasionally incorrect, producing stub records
that carry the canonical name and institution of an established
researcher but a mismatched ORCID, and therefore fail the automatic
merger. We audited prolific same-name groups for these splits and
surfaced $164$ candidate cases. The $14$ highest-confidence cases
have been resolved manually; the remaining $150$ are held back
pending stronger evidence, because at looser thresholds genuine
homonyms at shared institutions become indistinguishable from real
splits. A worst-case merge of every audited pair would add at most
$\approx 123$ coauthor edges, an upper bound of $0.15\%$ on
split-identity noise over the $\num{83035}$ edges in the final
graph. Appendix~\ref{app:disambig} reports the full audit, including
an LLM-based sanity check on borderline pairs. The dominant residual
failure mode is homonymy among common East Asian surnames where no
ORCID is available, which can inflate apparent productivity for a
small number of names; this affects the long tail rather than the
top contributors of Section~\ref{sec:descriptive}, where spot checks
of the top-30 centrality ranks found no misattributed records.

\subsection{Citation count reconciliation}
\label{subsec:data-citations}

We reconcile citation counts by taking the elementwise maximum of the
OpenAlex \texttt{cited\_by\_count} field and the Crossref
\texttt{is-referenced-by-count} field. The two counts disagree often
enough to matter, and inspection of the disagreements suggests that the
gap reflects differing coverage of the citing corpus rather than
differing quality. OpenAlex tends to index a broader non-English long
tail and more preprint-to-journal chains, while Crossref occasionally
captures venue-to-venue references that OpenAlex has yet to ingest.
Treating the two as complementary rather than competing therefore
yields a tighter lower bound on true citation impact than either source
in isolation.

The split across the $\num{120323}$ papers is uneven. Roughly $59\%$ of
papers carry a higher OpenAlex count than Crossref count, about $7\%$
carry a higher Crossref count, and a further $25\%$ agree exactly,
with the balance accounted for by papers lacking a count on one side. The asymmetry toward OpenAlex is consistent with its
wider ingestion of secondary literature, whereas the Crossref-leading
subset concentrates on older papers whose citing articles predate
OpenAlex's retrospective backfill.

This design choice has direct implications for comparisons against
\citet{sun2017coauthorship}, who drew citation counts from Web of
Science as of early 2017. Our totals run materially higher for two
independent reasons. Nine additional years of citations have accrued
since their snapshot, and the union of OpenAlex and Crossref indexes a
broader non-English and non-Anglophone long tail than Web of Science did
at that time. The implication is that absolute citation magnitudes are
not directly comparable across the two studies. However, the rank
ordering of the top-30 central authors proves considerably more stable
across source choices than the absolute numbers, so centrality-based
conclusions from Section~\ref{sec:coauthor} carry across the
reconciliation choice. All hyperparameters governing ingest,
deduplication, paper embeddings, graph construction, Leiden community
detection, and the phantom-collaborator evaluation are consolidated in
Appendix~\ref{app:hyperparams}.


\section{Descriptive bibliometrics}
\label{sec:descriptive}

\begin{figure}[width=0.99\textwidth, pos = t]
  \centering
  \begin{subfigure}[t]{0.48\linewidth}
    \centering
    \includegraphics[width=\linewidth]{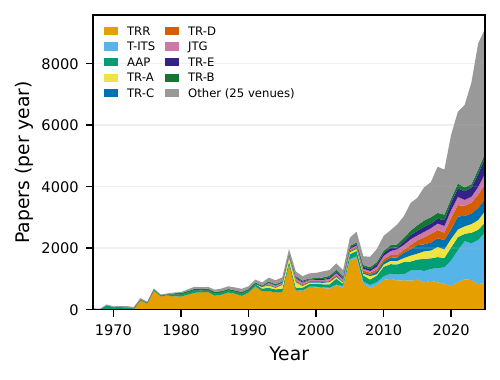}
    \caption{Papers per year.}
    \label{fig:papers-by-year}
  \end{subfigure}\hfill
  \begin{subfigure}[t]{0.48\linewidth}
    \centering
    \includegraphics[width=\linewidth]{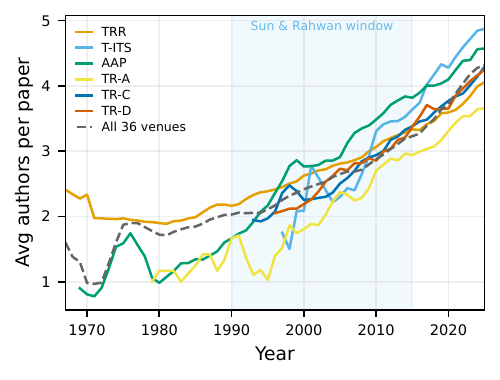}
    \caption{Team size over time.}
    \label{fig:team-size}
  \end{subfigure}\\[1ex]
  \begin{subfigure}[t]{0.48\linewidth}
    \centering
    \includegraphics[width=\linewidth]{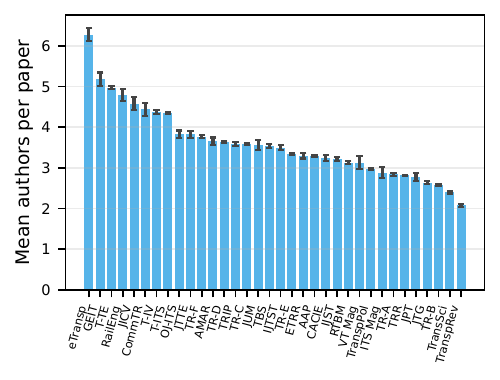}
    \caption{Team size by venue.}
    \label{fig:authors-per-venue-bar}
  \end{subfigure}
    \hfill
\begin{subfigure}[t]{0.48\linewidth}
  \centering
  \includegraphics[width=\linewidth]{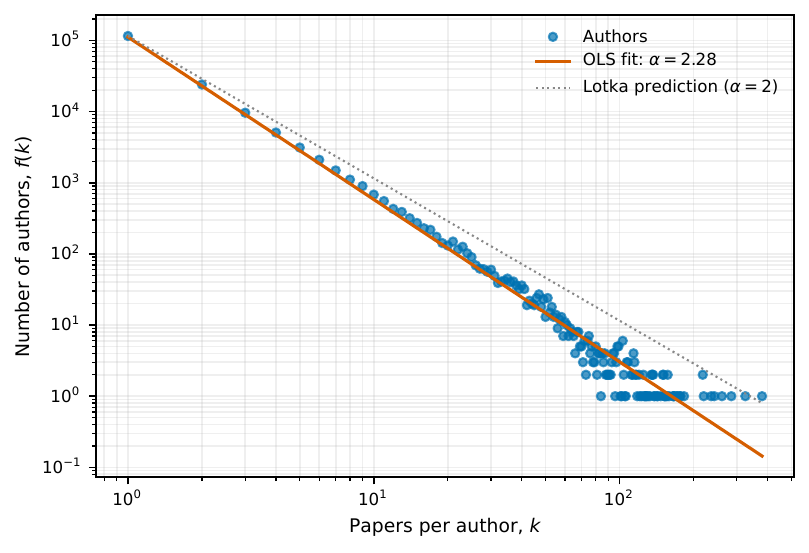}
  \caption{Author productivity distribution (log--log).}
  \label{fig:lotka}
\end{subfigure}
  \caption{Corpus growth, author-team size, and Lotka-style productivity.
  }
  \label{fig:growth-panel}
\end{figure}

\Cref{fig:growth-panel} situates the corpus along four dimensions that
together summarize how transportation research has scaled between
\citet{sun2017coauthorship} and the present. \Cref{fig:growth-panel}
(\subref{fig:papers-by-year}) plots annual paper counts stacked by venue,
so the height of a given band is the yearly output of that venue and
the total envelope is the field-wide annual volume. Transportation
Research Record (TRR) dominates the early decades and is effectively
the entire corpus before $1985$, because our OpenAlex coverage of the
other TR-family titles thins sharply as one moves back in time.
The Elsevier TR-family (Parts A--F) and the IEEE Transactions on
Intelligent Transportation Systems contribute the bulk of post-$2015$
growth, with T-ITS alone adding more than a thousand papers per year by
$2023$. Total annual volume surpasses ten thousand papers by the
mid-$2020$s, roughly an order of magnitude above the
$1990$--$2015$ window of S\&R and well above the volumes that other
transportation bibliometric reviews covered
(\citealp{jiang2020trb}; \citealp{modak2019fifty}).

\Cref{fig:growth-panel} (\subref{fig:team-size}) shows the three-year rolling mean of
authors per paper for the six highest-volume venues, with the
$1990$--$2015$ S\&R window shaded for reference. Every major venue
trends upward monotonically. TRR and the Elsevier TR-family rise
smoothly from a 1990s baseline of roughly two to three authors per
paper to a current level near four, consistent with the broader science
finding that team size drifts upward as data sharing, co-development,
and cross-institutional collaboration become cheaper
\citep{newman2001structure}. The IEEE venues (T-ITS and T-IV) accelerate
sharply after $2018$, reaching a mean above five authors per paper
and by $2024$ approaching six, as ML-heavy submissions with a
first-author student plus several co-advisors plus dataset contributors
arrive. The slope change is visible in the rolling mean and not merely
an artifact of a handful of large teams: the increase is accompanied by
a sharp drop in single-author fraction (reported in
Appendix~\ref{app:descriptive-extended}).

\Cref{fig:growth-panel} (\subref{fig:authors-per-venue-bar}) breaks the same quantity down
into a per-venue bar chart averaged over the full window, giving a
static ordering of venues by team-size preference. The IEEE-family
journals and the INFORMS operations-research outlet (Transportation
Science) lead on team size, reflecting the multi-author algorithmic and
engineering style typical of those communities. The behavioral-science
outlets (TR-F, Accident Analysis and Prevention) and the policy-oriented
TR-A sit at the lower end, consistent with the one-or-two-author
survey-and-analysis style that dominates their submissions. The ordering
is directly analogous to the finding of \citet{sun2017coauthorship} and
confirms that a decade of additional data has not reshuffled the
relative team-size ranking of the venues even as absolute team size has
drifted upward in every venue.

\Cref{fig:growth-panel} (\subref{fig:lotka}) reports the Lotka-style log--log frequency
distribution of author productivity: the $x$-axis is the number of
papers an author has in the corpus, the $y$-axis is the number of
authors with that exact count, and the dashed line is the OLS fit. The
fit gives exponent $\alpha \approx 2.28$ over $\num{168090}$ authors,
which sits between Lotka's \citeyearpar{lotka1926frequency} $\alpha = 2$
and the $\alpha \approx 2.6$ reported by
\citet{sun2017coauthorship}. The fitted slope is a power-law tail, not
a distinct mode: over 60\% of authors publish exactly one paper in the
corpus, and the long flat tail captures the handful of exceptionally
productive researchers that dominate the top-centrality tables of
Appendix~\ref{app:network-properties}. A flatter tail than S\&R
reported means our corpus captures more high-productivity authors in
absolute terms, consistent with the larger scale, while preserving the
same functional shape that Lotka first described for the physical
sciences a century ago.

Per-venue bibliometric totals (papers, unique authors, mean team size,
single-author fraction, mean citations, median career length) are
reported in Appendix~\ref{app:descriptive-extended}, along with the
top-five contributors and three most-cited papers for each venue.
These venue-level details serve as a reference but do not feed into the
coauthor-structure, topic-space, or phantom-collaborator results that
are the paper's central contributions.

\section{Coauthor network structure}
\label{sec:coauthor}

\subsection{Graph construction}
\label{subsec:coauthor-construction}

\begin{figure}[width=0.99\textwidth, pos = !t]
  \centering
  \begin{subfigure}[t]{0.42\linewidth}
    \centering
    \includegraphics[width=\linewidth]{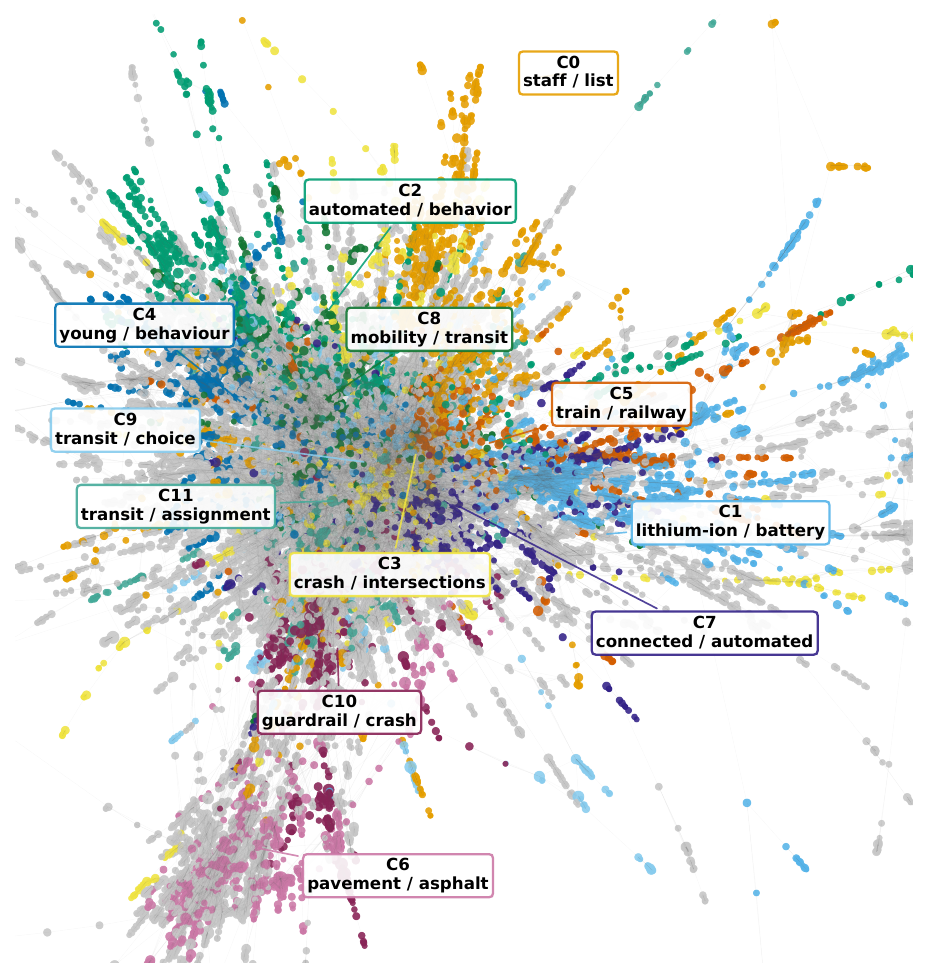}
    \caption{Community labels.}
    \label{fig:coauthor-overview}
  \end{subfigure}
  \begin{subfigure}[t]{0.48\linewidth}
    \centering
    \includegraphics[width=\linewidth]{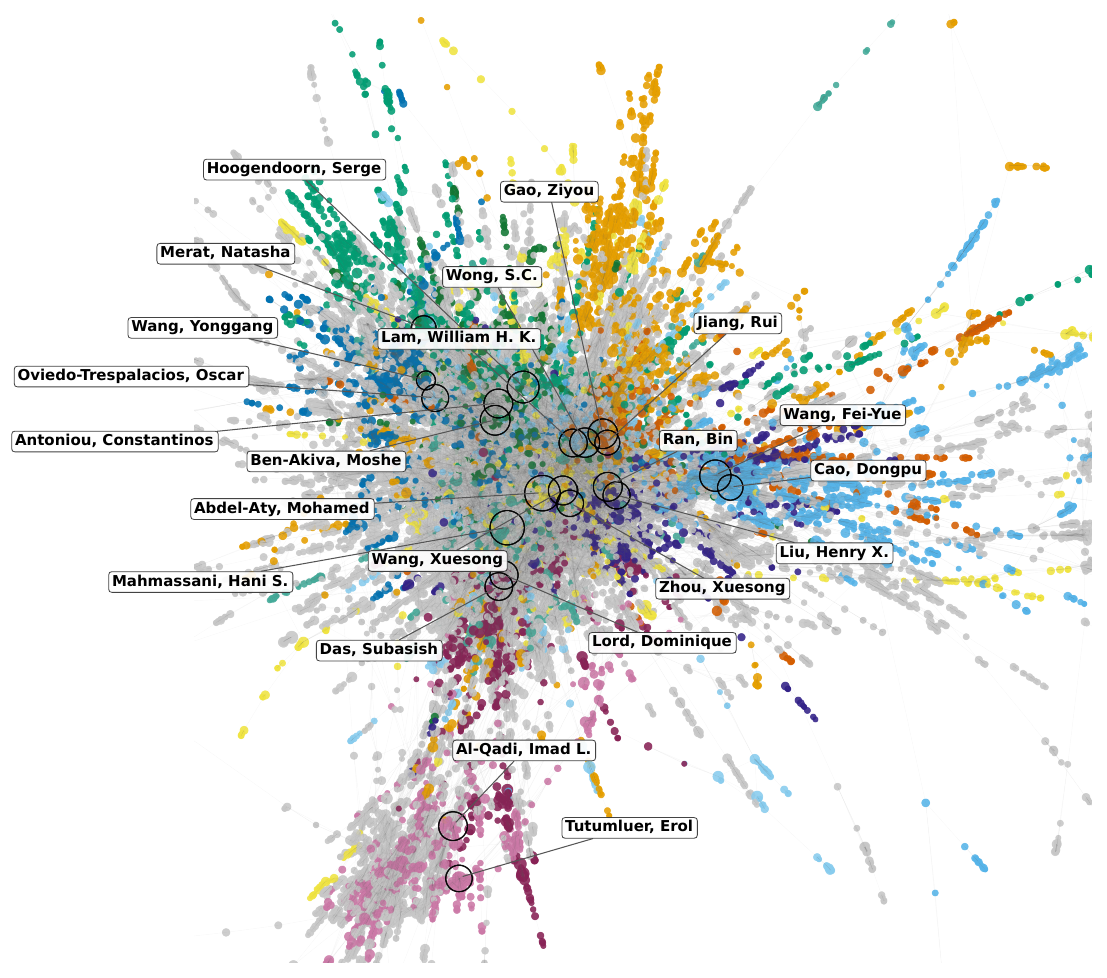}
    \caption{Top-degree author labels ($d \geq 55$).}
    \label{fig:lcc-labeled}
  \end{subfigure}
  \caption{Two views of the coauthor-network largest connected
    component (\num{28005} authors, top-12 Leiden
    communities colored).}
  \label{fig:coauthor-lcc}
\end{figure}

We build the coauthor graph from the deduplicated corpus by adding one
undirected edge for every pair of authors who share a paper. Authors are
canonicalized through the resolution chain described in
Sec.~\ref{sec:data} (OpenAlex ID, then ORCID, then a normalized surname
plus first-initial key), so a single researcher is represented by one node
even when individual papers disagree on romanisation or on the precise
first-name rendering. We then keep nodes with at least two papers in the
corpus so that the network reflects researchers who recurrently publish in
transportation. The resulting graph contains $\num{41551}$ author nodes
and $\num{83035}$ coauthor edges. We retain an edge in the displayed graph
when the two authors have coauthored at least twice, which is the same
threshold used in \citet{sun2017coauthorship}. For completeness, every
node additionally carries a list of all of its single-collaboration
partners so that reviewers can verify pairs that fall below the display
threshold without re-running the pipeline. \Cref{fig:coauthor-overview}
shows the full graph laid out by the precomputed ForceAtlas2 layout
\citep{jacomy2014forceatlas2}, with the 12 largest Leiden communities
colored and node size scaled by paper count.

\Cref{fig:coauthor-lcc} shows two views of the coauthor network's
largest connected component. Both panels share the same ForceAtlas2
layout, the same top-12 Leiden community coloring, and the same
axes crop --- only the label layer differs. Panel
\subref{fig:coauthor-overview} annotates the $12$ colored regions
with their Leiden community id C0--C11 and the two keyword-label
tokens inferred from paper titles inside each community.
Panel \subref{fig:lcc-labeled} annotates the same graph with a
geographically spread sample of high-degree authors, mirroring
\citet{sun2017coauthorship}. The LCC contains \num{28005}
authors. Familiar names anchor these communities: established
travel-demand and traffic-engineering figures (Ben-Akiva, Lam,
Mahmassani, Hoogendoorn, Wong) appear alongside safety-research
leaders (Abdel-Aty, Lord) and more recently central names from
automated-driving, pavement-infrastructure, and driver-behavior
research (Wang~Fei-Yue, Cao~Dongpu, Al-Qadi, Tutumluer, Merat).

\begin{figure}[width=0.99\textwidth, pos = t]
  \centering
  \includegraphics[width=0.54\linewidth]{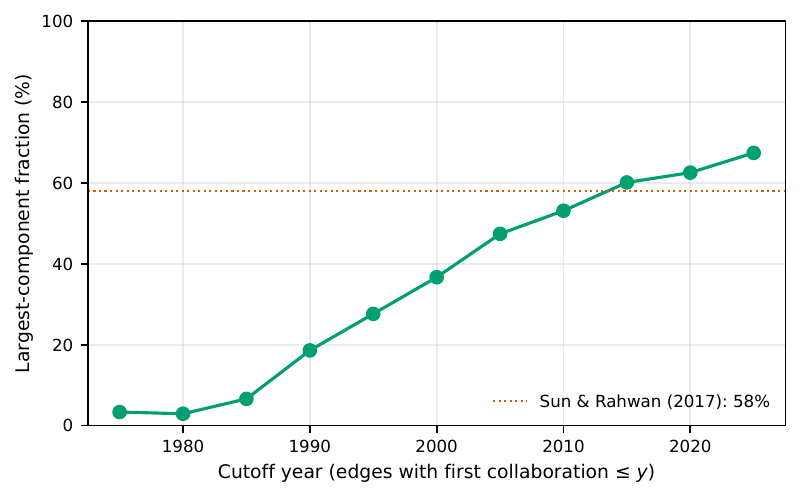}
  \caption{Giant-component fraction versus cutoff year.}
  \label{fig:giant-component}
\end{figure}

\Cref{fig:giant-component} plots the fraction of authors inside the
largest connected component as a function of the cutoff year, for the
full displayed graph ($d \geq 1$ collaborations in the window). The
giant component contained only $3.4\%$ of authors in $1975$ but grows
monotonically to $\mathbf{67.4\%}$ of the \num{41551} authors in the
displayed graph by $2025$, markedly above the $58\%$ that
\citet{sun2017coauthorship} reported for their 1990--2015 TR-family
subset. The $9.4$ percentage-point gap reflects two factors. First,
our 34-venue corpus spans both the Elsevier TR-family and the IEEE
intelligent-transportation venues, where cross-venue authorship is
now common. Second, the additional decade of data since 2015 captures
the sharp rise of multi-institution ML collaborations that pushes
previously isolated clusters into the giant component. In short,
transportation research has crossed the connectivity threshold that
\citet{newman2001structure} argued characterizes a mature scientific
community. Standard network-science diagnostics (degree, strength,
citation, and shortest-path distributions; centrality correlations;
and top-$30$ authors by combined centrality) appear in
Appendix~\ref{app:network-properties}.

\subsection{Leiden communities}
\label{subsec:coauthor-leiden}

We partition the coauthor graph with the Leiden algorithm
\citep{traag2019leiden} using an RB-configuration vertex partition, as
recommended by \citet{fortunato2016community}. We merge singleton or
under-sized components ($<$10 members) into a single miscellaneous bucket,
leaving $\mathbf{172}$ mainland communities. This is roughly eight times
the $\sim$20 communities reported by \citet{sun2017coauthorship}, which is
expected given our larger corpus and finer resolution parameter.

\Cref{tab:top-communities} lists the twenty largest communities with their
keyword labels (top-TF-IDF words from the titles of the community's papers)
and three exemplar authors per community. The labels reveal a clear
topical structure. The largest communities are associated with freight
logistics and supply-chain modeling, traffic-flow theory, travel-demand
choice modeling, pavement and materials engineering, and, more recently,
connected and automated vehicles. These groupings broadly match the
editorial scopes of TR Parts E, B, A, D, and C, respectively.

\begin{table}[width=0.99\textwidth, pos=t]
  \centering
  \small
  \caption{Twenty largest coauthor-Leiden communities.}
  \label{tab:top-communities}
  \resizebox{\textwidth}{!}{
  \begin{tabular}{rlll}
  \toprule
  \textbf{\#} & \textbf{Size} & \textbf{Keyword label} & \textbf{Exemplar authors} \\
  \midrule
  0 & 1,891 & staff, list, routing, shipping, container, supply & Irvine, James, Gozalvez, Javier, Zhuang, Weihua, Uhleman, Elisabeth, Melley, Dawn \\
  1 & 1,629 & lithium-ion, battery, batteries, trajectory, energy, strategy & Wang, Fei-Yue, Li, Li, Cao, Dongpu, Lv, Chen, Cai, Yingfeng \\
  2 & 1,277 & automated, behavior, behaviour, effects, public, choice & Hoogendoorn, Serge, Merat, Natasha, Arem, Bart Van, Cats, Oded, Lint, Hans Van \\
  3 & 1,129 & crash, intersections, crashes, behavior, risk, pedestrian & Abdel-Aty, Mohamed, Wang, Xuesong, Huang, Helai, Lee, Jaeyoung, Quddus, Mohammed \\
  4 & 949 & young, behaviour, use, crash, risk, effects & Oviedo-Trespalacios, Oscar, Haque, Md. Mazharul, Sun, Jian, Washington, Simon, Zheng, Zuduo \\
  5 & 940 & train, railway, rail, metro, urban rail, transit & Gao, Ziyou, Tang, Tao, Yang, Lixing, Jiang, Rui, Wu, Jianjun \\
  6 & 882 & pavement, asphalt, concrete, pavements, mixtures, asphalt mixtures & Al-Qadi, Imad L., Tutumluer, Erol, Darter, M I, Flintsch, Gerardo W., Underwood, B. Shane \\
  7 & 863 & connected, automated, cooperative, connected automated, trajectory, flow & Ran, Bin, Barth, Matthew, Liu, Henry X., Peeta, Srinivas, Yin, Guodong \\
  8 & 827 & mobility, transit, public, choice, demand, behavior & Ben-Akiva, Moshe, Antoniou, Constantinos, Yannis, George, Koutsopoulos, Haris N., Zhao, Jinhua \\
  9 & 809 & transit, choice, public, effects, demand, behavior & Wong, S.C., Lam, William H. K., Szeto, W.Y., Li, Zhichun, Sumalee, Agachai \\
  10 & 738 & guardrail, crash, crashes, bridge, intersections, bridge rail & Das, Subasish, Faller, Ronald K., Lord, Dominique, Zhao, Xiaohua, Zhang, Yunlong \\
  11 & 727 & transit, assignment, demand, charging, routing, freight & Mahmassani, Hani S., Zhou, Xuesong, Nie, Yu, Zockaie, Ali, Saberi, Meead \\
  12 & 712 & transit, accessibility, public, canada, use, mobility & Witlox, Frank, El-Geneidy, Ahmed, Cheng, Long, Vos, Jonas De, Ettema, Dick \\
  13 & 701 & concrete, pavement, asphalt, freight, pavements, choice & Ozbay, Kaan, Holguin-Veras, Jose, Harvey, John, Yang, Hong, Xie, Kun \\
  14 & 682 & asphalt, pavement, mixtures, concrete, pavements, asphalt mixtures & Little, Dallas N., Lytton, Robert L., Choubane, Bouzid, Wang, Kelvin C. P., Bahia, Hussain U. \\
  15 & 662 & pavement, concrete, pavements, evacuation, transit, choice & Waller, S. Travis, Mohammadian, Abolfazl, Ukkusuri, Satish V., Dixit, Vinayak, Rashidi, Taha Hossein \\
  16 & 658 & crash, flux, pmsm, effects, behavior, machine & Liu, Pan, Sayed, Tarek, Wang, Wei, Sze, N.N., Li, Zhibin \\
  17 & 655 & crash, behavior, automated, pedestrian, effects, safety climate & Noyce, David A., Fisher, Donald L., Knodler, Michael, Chitturi, Madhav, Ritchie, Stephen G. \\
  18 & 607 & ride-sourcing, equilibrium, pricing, services, demand, transit & Yang, Hai, Chen, Xiqun, Schonfeld, Paul, Yin, Yafeng, Li, Meng \\
  19 & 532 & choice, public, mobility, australia, behaviour, use & Hensher, David A., Nelson, John D., Hess, Stephane, Polak, John, Mulley, Corinne \\
  \bottomrule
  \end{tabular}
  }
\end{table}

\subsection{Bridge edges between communities}
\label{subsec:coauthor-bridges}

\begin{figure}[width=0.99\textwidth, pos = !t]
  \centering
  \includegraphics[width=0.82\linewidth]{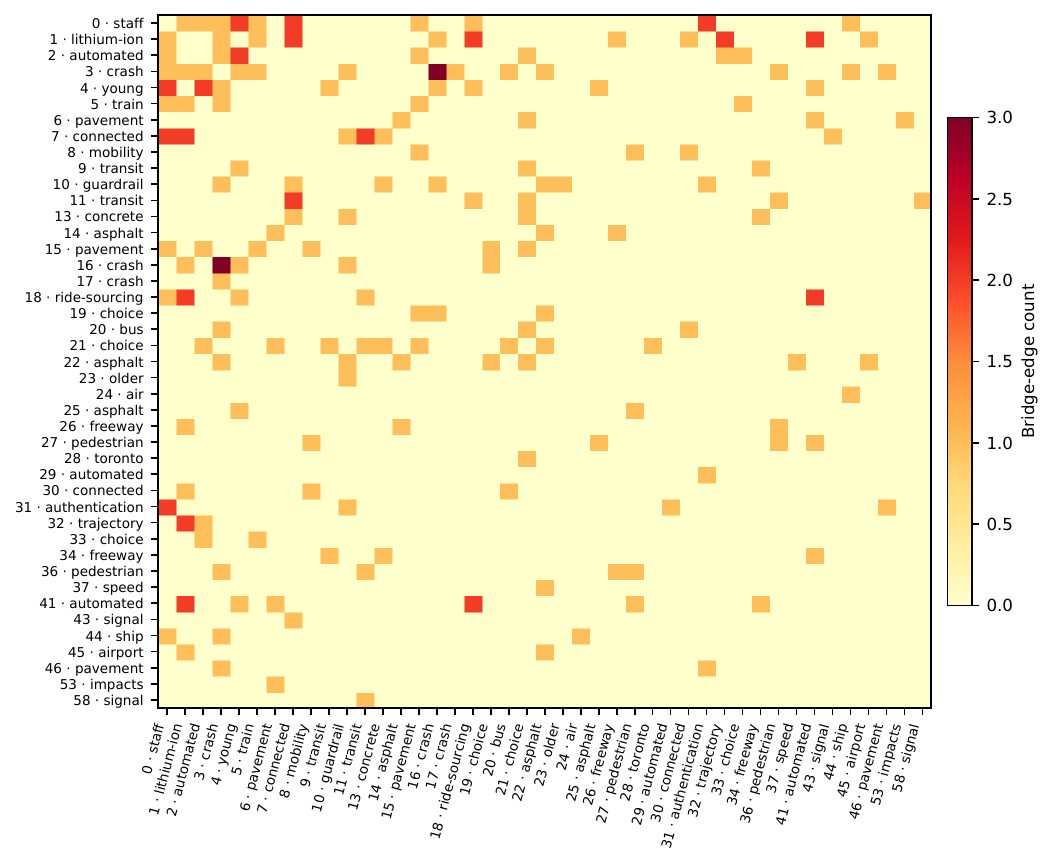}
  \caption{Top-$100$ bridge edges, aggregated by community pair.}
  \label{fig:bridges}
\end{figure}

\Cref{fig:bridges} visualizes the top-$100$ bridge edges, defined as the
edges with the highest edge-betweenness whose endpoints sit in different
Leiden communities. Because bridge-edge betweenness is proportional to
the number of shortest paths passing through the edge, these are the ties
that do the most work holding the network together. The figure shows
that the bridges are not uniform: a small number of community pairs
account for most of the brokerage, notably traffic-flow to CAV,
optimization to behavior, and freight to intermodal logistics. These are
exactly the topical interfaces where the phantom-collaborator test of
Sec.~\ref{sec:phantom} has the most purchase, because they are both
(i)~sparsely coauthored in the training window and (ii)~semantically
close enough that hybrid-embedding cosine picks up candidate bridges.

Taken together, Secs.~\ref{sec:descriptive} and \ref{sec:coauthor}
\textbf{answer RQ1}: transportation research has scaled by roughly an
order of magnitude in paper volume since \citet{sun2017coauthorship}, the
giant component has grown from $58\%$ to $67.4\%$, the number of Leiden
communities has grown from $\sim 20$ to $172$, and the Lotka exponent
has flattened from $\alpha \approx 2.6$ to $\alpha \approx 2.28$ --- all
consistent with the \citet{newman2001structure} characterization of a
mature, tightly connected scientific community.

\section{Semantic structure}
\label{sec:semantic}

\subsection{Author centroids and UMAP projection}
\label{subsec:semantic-centroids}

We embed every paper $p$ with SPECTER2
\citep{cohan2020specter, singh2023specter2}, a citation-aware pretrained
language model that produces a vector
$\mathbf{s}_p \in \mathbb{R}^{768}$ from the title and abstract using a
SciBERT backbone fine-tuned on a contrastive citation-pair objective.
We aggregate to the author level by a citation-weighted mean, then
$L_2$-normalize:
\begin{equation}
  \bar{\mathbf{s}}_i
  \;=\; \frac{\sum_{p \in P_i} w_p \, \tilde{\mathbf{s}}_p}
             {\bigl\lVert \sum_{p \in P_i} w_p \, \tilde{\mathbf{s}}_p \bigr\rVert_2},
  \qquad w_p \;=\; 1 + \log(1 + c_p),
  \label{eq:author-centroid}
\end{equation}
where $P_i$ is author $i$'s paper set, $c_p$ is $p$'s citation count,
and $\tilde{\mathbf{s}}_p$ is the whitened SPECTER2 vector
(Sec.~\ref{subsec:semantic-whitening}). The log-citation weight
dampens the contribution of papers that attracted little readership
without discarding them.

To sharpen local structure we concatenate three $L_2$-normalized
components with square-root weights:
\begin{equation}
  \mathbf{h}_i
  \;=\;
  \bigl[\, \sqrt{w_s}\,\bar{\mathbf{s}}_i
     \;\Vert\; \sqrt{w_c}\,\mathbf{c}_i
     \;\Vert\; \sqrt{w_v}\,\mathbf{v}_i \,\bigr],
  \qquad (w_s, w_c, w_v) = (0.55,\, 0.30,\, 0.15),
  \label{eq:hybrid-embedding}
\end{equation}
where $\mathbf{c}_i \in \mathbb{R}^{128}$ is a TF-IDF projection of
author $i$'s OpenAlex concepts (levels $\geq 2$) and
$\mathbf{v}_i \in \mathbb{R}^{28}$ is a venue-LDA projection; $\Vert$
denotes concatenation. The square-root scaling is what makes the
mixing weights interpretable: because each component is unit-length,
the pairwise cosine similarity in the hybrid space decomposes exactly as
\begin{equation}
  \cos(\mathbf{h}_i, \mathbf{h}_j)
  \;=\;
  w_s\, \cos(\bar{\mathbf{s}}_i, \bar{\mathbf{s}}_j)
  \;+\; w_c\, \cos(\mathbf{c}_i, \mathbf{c}_j)
  \;+\; w_v\, \cos(\mathbf{v}_i, \mathbf{v}_j),
  \label{eq:cosine-decomposition}
\end{equation}
so $(w_s, w_c, w_v)$ are literally the share that each signal
contributes to any downstream similarity computation. SPECTER2
dominates, with supervised concept and venue structure preserved at a
controlled contribution that would otherwise be compressed by the
dense mono-domain SPECTER2 geometry.

\begin{figure}[width=0.99\textwidth, pos = t]
  \centering
  \includegraphics[width=0.82\linewidth]{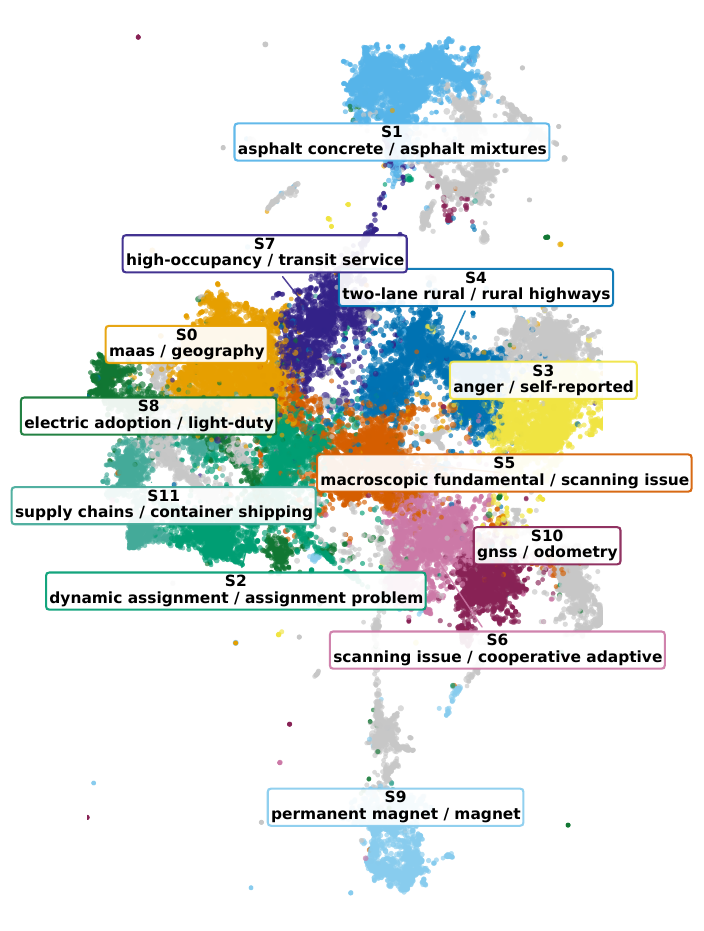}
  \caption{Author topic space (UMAP projection of the hybrid embedding),
    colored by semantic-Leiden community.}
  \label{fig:topic-space}
\end{figure}

\Cref{fig:topic-space} is a 2D projection of the author hybrid space
computed with UMAP
\citep[Uniform Manifold Approximation and Projection,][]{mcinnes2018umap}.
Colors are semantic-Leiden community IDs
(Sec.~\ref{subsec:semantic-leiden}). Clusters are visually separated
despite no community signal being passed to UMAP. The three macro-regions
that emerge correspond to \textbf{travel demand and behavior} (left),
\textbf{operations and optimization} (center), and \textbf{vehicle
autonomy and perception} (right), matching the disciplinary boundaries
used by the editorial offices of the TR-family journals.

\subsection{Whitening and its effect on pairwise cosine}
\label{subsec:semantic-whitening}

\begin{figure}[width=0.99\textwidth, pos = t]
  \centering
  \includegraphics[width=0.65\linewidth]{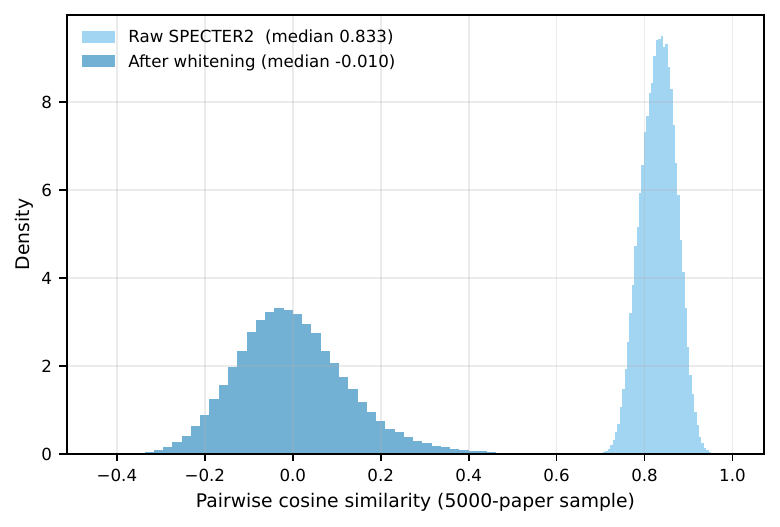}
  \caption{Pairwise cosine similarity, before and after
    \citet{arora2017simple} whitening.}
  \label{fig:whitening}
\end{figure}

A well-known property of transformer-based sentence embeddings is
\textbf{anisotropy}. All vectors share a dominant direction that encodes
generic document features, compressing the angular distance between any
two documents toward a narrow band. For our corpus, the median pairwise
cosine between two arbitrary paper embeddings is $\mathbf{0.833}$, meaning
almost every paper ``looks similar'' to every other paper. This is fatal
for any kNN-style downstream task. We apply the ``all-but-the-top''
correction of \citet{arora2017simple}. We subtract the mean, project out
the top-$1$ principal direction, and $z$-score each dimension. The
correction shifts the median pairwise cosine from $0.833$ to
$-0.010$ (Fig.~\ref{fig:whitening}). The top-$1$ principal
direction explains $9.6\%$ of the variance, yet its removal
is the decisive step because the direction encodes the corpus-level mean.
Once removed, the geometry becomes discriminative and the semantic-kNN
structure required by Sec.~\ref{sec:phantom} becomes meaningful.

\subsection{Semantic-Leiden communities}
\label{subsec:semantic-leiden}

To partition the author hybrid space into semantic communities we build a
mutual $k$-nearest-neighbor graph ($k=20$) and apply Leiden with the RB
configuration partition at resolution $1.0$. Small islands are collapsed
into a single miscellaneous bucket, leaving $\mathbf{23}$ mainland
semantic communities. \Cref{tab:semantic-communities} lists the top
communities with their keyword labels and exemplar authors. The labels
match recognizable subfields of transportation, for instance
\textbf{pavement materials}, \textbf{traffic-flow theory},
\textbf{activity-based travel demand}, \textbf{connected and automated
vehicles}, \textbf{road safety}, \textbf{freight and logistics}, and
\textbf{airline operations}. Unlike the coauthor Leiden partition, which
reflects social ties, this partition reflects topical writing style.
Sec.~\ref{sec:multiplex} compares the two.

\begin{table}[width=0.99\textwidth, pos=t]
  \centering
  \small
  \caption{Largest semantic-Leiden communities.}
  \label{tab:semantic-communities}
  \resizebox{\textwidth}{!}{
    \begin{tabular}{rlll}
  \toprule
  \textbf{\#} & \textbf{Size} & \textbf{Keyword label} & \textbf{Exemplar authors} \\
  \midrule
  0 & 6,426 & maas, geography, stated preference, car use, well-being, bicycling & Hensher, David A., Bhat, Chandra R., Timmermans, Harry, Ben-Akiva, Moshe, Axhausen, Kay W. \\
  1 & 4,180 & assignment problem, problem windows, location problem, berth, pickup delivery, liner shipping & Mahmassani, Hani S., Yang, Hai, Meng, Qiang, Wang, Shuaian, Gao, Ziyou \\
  2 & 3,859 & anger, novice, hazard perception, self-reported, sleep, phone use & Wets, Geert, Oviedo-Trespalacios, Oscar, Brijs, Tom, Yan, Xuedong, Wong, Yiik Diew \\
  3 & 3,203 & asphalt concrete, asphalt mixtures, hot-mix, hot-mix asphalt, concrete pavement, concrete pavements & Al-Qadi, Imad L., Sinha, Kumares C., Harvey, John, Lytton, Robert L., Darter, M I \\
  4 & 3,052 & actuator, fault-tolerant, finite-time, mpc, path following, sliding mode & Li, Li, Cao, Dongpu, Lv, Chen, Li, Shengbo Eben, Li, Keqiang \\
  5 & 2,855 & light-duty, plug-in hybrid, plug-in electric, energy use, fast charging, fuel cell & Barth, Matthew, Yu, Lei, Guensler, Randall, Song, Guohua, Niemeier, Deb \\
  6 & 2,842 & sight distance, two-lane rural, rural highways, safety functions, rural two-lane, injury severities & Abdel-Aty, Mohamed, Sayed, Tarek, Khattak, Asad J., Wang, Xuesong, Lord, Dominique \\
  7 & 2,758 & high-occupancy, commuter rail, small urban, planning process, trip generation, san diego & Shalaby, Amer, Furth, Peter G., Levinson, Herbert S., Hickman, Mark, Machemehl, Randy B. \\
  8 & 2,666 & scanning issue, gnss, odometry, slam, stereo, pedestrian detection & Wang, Fei-Yue, Eskandarian, Azim, Sotelo, Miguel Angel, Chen, Long, Wang, Xiao \\
  9 & 2,647 & highway capacity, macroscopic fundamental, capacity manual, scanning issue, continuum, perimeter & Hoogendoorn, Serge, Wong, S.C., Rakha, Hesham, Rouphail, Nagui M., Ran, Bin \\
  10 & 2,523 & soils, cement, decks, concrete bridge, pipe, guardrail & Tutumluer, Erol, Faller, Ronald K., Sicking, Dean L., Puppala, Anand J., Bligh, Roger P \\
  11 & 2,510 & permanent magnet, pmsm, converter, voltage, torque, flux & Hua, Wei, Emadi, Ali, He, Zhengyou, Sun, Xiaodong, Zhang, Zhuoran \\
  12 & 2,265 & supply chains, container shipping, e-commerce, hinterland, chain management, liner shipping & Holguin-Veras, Jose, Sheu, Jiuh-Biing, Yang, Zaili, Macharis, Cathy, Tavasszy, Lorant \\
  13 & 2,231 & alcohol, fatalities, child, helmet, accid, prev & Elvik, Rune, Yannis, George, Hauer, Ezra, Kim, Karl, Haworth, Narelle \\
  14 & 2,059 & authentication, vanets, privacy-preserving, federated learning, president, blockchain-based & Gozalvez, Javier, Glickenstein, Harvey, Vlacic, Ljubo, Olaverri-Monreal, Cristina, Yu, F. Richard \\
  15 & 1,578 & airlines, airline industry, domestic, evidence china, air cargo, air flow & Zhang, Anming, Fu, Xiaowen, Hansen, Mark, Wang, Kun, Jiang, Changmin \\
  16 & 1,365 & short-term flow, mml, editor column, research lab, federated learning, graph convolution & Wang, Yinhai, Antoniou, Constantinos, Chen, Xiqun, Lv, Yisheng, Qian, Sean \\
  17 & 1,236 & pavement crack, damage detection, reinforced concrete, crack segmentation, finite element, building structures & Adeli, Hojjat, Tsai, Yichang, Wang, Kelvin C. P., Qin, Yong, Jia, Limin \\
  18 & 1,169 & lithium-ion batteries, fuel cell, estimation lithium-ion, state-of-health, health estimation, energy storage & Wang, Zhenpo, Ouyang, Minggao, Tang, Xiaolin, Wei, Zhongbao, Zhou, Quan \\
  19 & 551 & ballast, railway track, alignment optimization, wheel rail, derailment, railway alignment & Schonfeld, Paul, Barkan, Christopher P. L., Dick, C. Tyler, Lai, Yung-Cheng, Qian, Yu \\
  20 & 387 & hurricane evacuation, evacuations, wildfire, during hurricane, hurricanes, infrastructure resilience & Wolshon, Brian, Murray-Tuite, Pamela, Du, Bo, Wilmot, Chester G., Hasan, Samiul \\
  21 & 249 & calendar, list, adeli, scanning issue, book reviews, editor column & Lauer, Martin, Gozalvez, Javier, Wang, Dawei, Beck, J, Daniel, D \\
  22 & 209 & highway noise, noise exposure, noise barrier, noise barriers, noise abatement, annoyance & Donavan, Paul R., Cai, Ming, Wayson, Roger L., Cohn, Louis F., Bowlby, William \\
  \bottomrule
  \end{tabular}
  }
\end{table}

\subsection{Why the production embedding is the hybrid}
\label{subsec:semantic-ablation}

We ship the three-component hybrid (whitened SPECTER2 plus concept
TF-IDF plus venue LDA) rather than raw SPECTER2, even though a small
ablation we ran on this corpus gives raw a roughly one- to
two-percentage-point edge on phantom-collaborator precision at $K = 20$.
The reason is that the phantom test measures only top-$K$ nearest-neighbor
retrieval, which is robust to the anisotropy that otherwise makes raw
SPECTER2 nearly unusable: the rank order of the closest few neighbors is
preserved even when the median pairwise cosine is $0.83$ and almost
every pair looks similar in absolute terms. Every other downstream
artifact in this paper depends on a discriminative, well-spread
geometry. The UMAP projection in Sec.~\ref{subsec:semantic-centroids},
the mutual-kNN semantic Leiden of Sec.~\ref{subsec:semantic-leiden}, the
multiplex construction of Sec.~\ref{sec:multiplex}, and the trajectory
taxonomy of Sec.~\ref{sec:trajectories} all require that typical pairs
sit near zero similarity so that the rare genuinely-close pairs are
visible. Without whitening those four artifacts collapse into a single
blob. Adding the concept-TF-IDF and venue-LDA layers injects supervised
topical signal that the pure sentence embedding misses and recovers
most of the retrieval gap while preserving the community structure.
Shipping the hybrid therefore trades at most two percentage points of
phantom-collaborator precision for a geometry that supports every other
analysis in the paper.

\section{Multiplex communities: where coauthorship and topic diverge}
\label{sec:multiplex}

\subsection{Multiplex graph construction}
\label{subsec:multiplex-construction}

Coauthorship encodes who has already worked with whom, whereas semantic
similarity encodes who writes about similar problems. These two signals
need not agree. A pair of authors in the same topical niche may belong to
different labs, institutions, or countries and therefore never coauthor,
while a pair of close collaborators may drift into divergent topics over
time. We examine the discrepancy with a multiplex graph that combines
both layers, following the formalism of \citet{mucha2010community}.

We construct the multiplex graph on the $\num{41551}$ authors of the
coauthor graph (Sec.~\ref{sec:coauthor}). The coauthor layer contributes a
weighted edge $(a, b)$ with weight $\alpha \log(1 + w_{ab})$, where
$w_{ab}$ is the number of papers that $a$ and $b$ have coauthored and
$\alpha = 0.5$. The semantic layer contributes an edge whenever both $a$
and $b$ list each other among their top-$5$ semantic neighbors in the
hybrid space (mutual-kNN) and the pairwise cosine similarity exceeds
$\tau_s = 0.60$. The semantic edge weight is
$(1-\alpha)(\cos(a, b) - \tau_s) / (1 - \tau_s)$, a soft rescaling that
leaves the strongest semantic ties comparable in magnitude to the
strongest coauthor ties. Running Leiden at resolution $0.5$ on the merged
graph and collapsing islands yields $\mathbf{181}$ combined
communities, larger than both input partitions.

\subsection{Partition alignment}
\label{subsec:multiplex-alignment}

We compare the three partitions with three standard measures. Normalised
mutual information (NMI) captures how predictable one partition is from
another, scaled to $[0, 1]$. Adjusted Rand index (ARI) measures pair-level
agreement corrected for chance. Variation of information (VI) is the
Shannon-information distance between partitions, measured in bits, and is
unbounded above. \Cref{tab:partition-alignment} reports all three.

\begin{table}[width=0.99\textwidth, pos = t]
  \centering
  \small
  \caption{Pairwise alignment between the three Leiden partitions
    (NMI, ARI, VI).}
  \label{tab:partition-alignment}
    \begin{tabular}{llrrrrr}
  \toprule
  \textbf{A} & \textbf{B} & \textbf{NMI ($\uparrow$)} & \textbf{ARI ($\uparrow$)} & \textbf{VI bits ($\downarrow$)} & \textbf{$|A|$} & \textbf{$|B|$} \\
  \midrule
  Coauthor & Semantic & 0.234 & 0.044 & 7.50 & 173 & 23 \\
  Coauthor & Combined & 0.463 & 0.093 & 5.88 & 173 & 180 \\
  Semantic & Combined & 0.363 & 0.164 & 6.07 & 23 & 180 \\
  \bottomrule
  \end{tabular}
\end{table}

Two observations are notable. First, the coauthor and semantic partitions
are only weakly aligned (NMI $=0.23$, ARI $=0.04$). This low alignment is
the \emph{raison d'\^etre} of the phantom-collaborator test in
Sec.~\ref{sec:phantom}. If coauthorship and topic already coincided, there
would be no hidden pairs to predict. Second, the combined partition
recovers roughly half of the coauthor information (NMI $=0.46$) and about
one-third of the semantic information (NMI $=0.36$). In other words, the
multiplex partition borrows more structure from the coauthor layer than
from the semantic layer, but not by a large margin. Both layers
contribute.

\begin{figure}[width=0.99\textwidth, pos = t]
  \centering
  \includegraphics[width=0.85\linewidth]{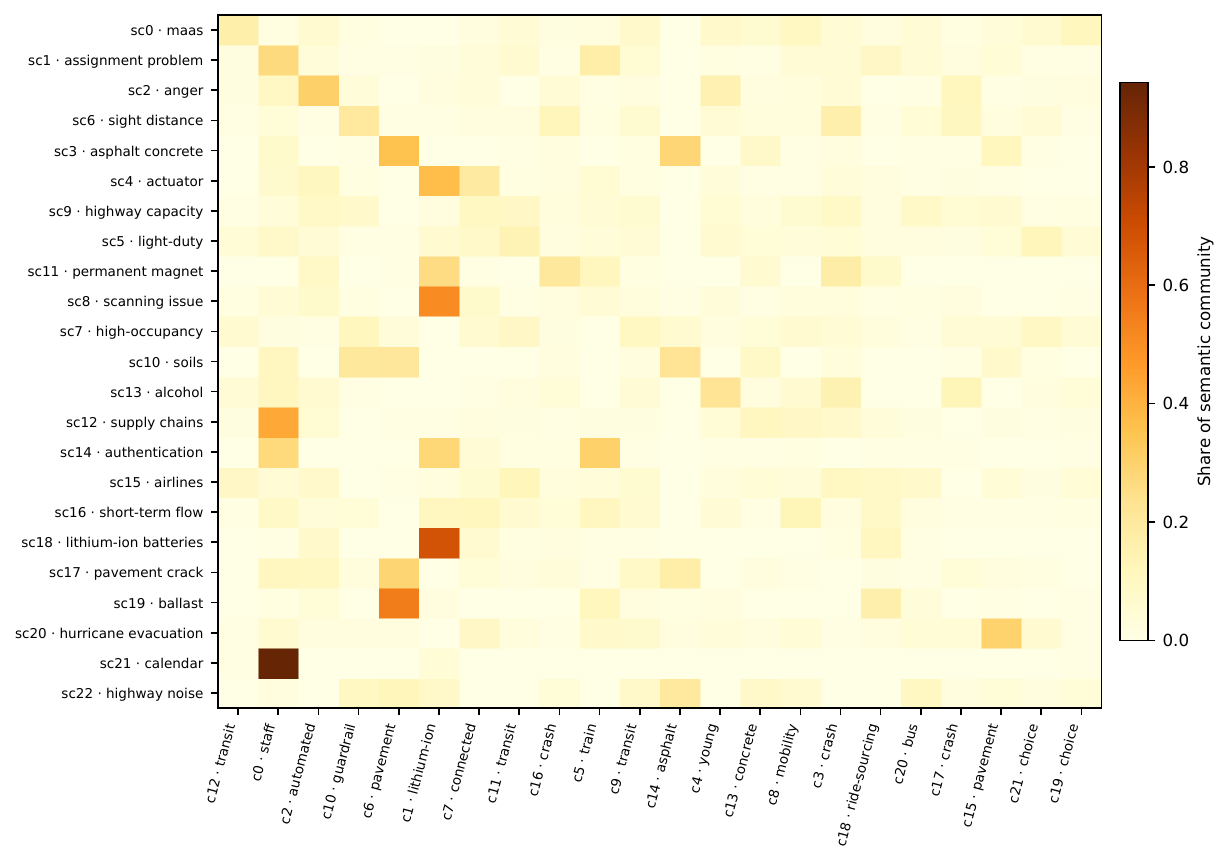}
  \caption{Row-normalized co-occurrence between semantic communities
    (rows) and top-$22$ coauthor communities (columns).}
  \label{fig:partition-cooc}
\end{figure}

\Cref{fig:partition-cooc} visualizes the row-normalized co-occurrence
matrix between the top-$22$ coauthor communities and all semantic
communities. A single semantic community typically spreads across several
coauthor communities, confirming that researchers writing similar papers
often sit in socially disjoint components of the collaboration graph.

The NMI $=0.23$ is the quantitative \textbf{answer to RQ2}: topic and
coauthorship structure agree only weakly. This is what makes a predictive
layer possible rather than redundant; if the two partitions already
coincided, the phantom-collaborator test of Sec.~\ref{sec:phantom} would
reduce to a recovery of structure the coauthor graph already encodes.

\subsection{Interpreting the divergence}
\label{subsec:multiplex-cases}

The weak coauthor--semantic alignment has three mechanistic drivers worth
naming. First, geography segregates researchers who study the same
problem. Pavement-materials research, for example, is one of the tightest
semantic communities in our data, yet the major groups (North American,
European, Chinese, and Middle Eastern) have limited mutual coauthorship.
Second, editorial scope drives venue-specific writing conventions even
when topics overlap. Choice modelers publishing in TR-B and in
Transportation Science pick different notation and emphasis, which the
hybrid embedding picks up. Third, generational turnover means that senior
authors who defined a subfield rarely appear as coauthors on the current
wave of ML-driven follow-up papers, even when both generations occupy the
same semantic community. The multiplex graph makes all three drivers
observable in a single partition, which is the motivation for using it as
the substrate of the static atlas described in Sec.~\ref{sec:atlas}.

\section{Phantom collaborators: a predictive test}
\label{sec:phantom}

\subsection{Why a predictive test, and its ceiling}
\label{subsec:phantom-motivation}

The analyses in Secs.~\ref{sec:descriptive}--\ref{sec:multiplex} describe
the current shape of the network; none of them takes a position on where
the field will go next. A forward-looking counterpart matters because
transportation research increasingly depends on collaborations that
existing social-graph-based recommender systems systematically miss: the
necessary ties have not yet formed, and classical link-prediction
baselines such as popularity or shared venue default to proposing authors
who are already central. An open, semantic atlas is the natural place to
test whether paper-embedding proximity carries collaboration-relevant
information that coauthorship structure alone does not encode. If it
does, the finding is both a validation of the hybrid embedding built in
Sec.~\ref{sec:semantic} and the basis for a candidate-partner feature
inside the live atlas (Sec.~\ref{sec:atlas}).

Coauthorship, however, is a rare event whose formation depends on
signals that lag or sit outside the published record. A doctoral
student's nascent research interests may precede their published
output by several years; funding-driven team composition, institutional
hiring, and in-person conference conversations are all likewise
invisible to a text-only embedding. A text-based predictor can surface
pairs whose published centroids have already converged but whose
social graph has not yet caught up; it cannot surface pairs whose
future topical alignment is still forming and not yet visible in the
corpus, nor pairs whose collaboration will form through a channel the
corpus does not capture. This is an intrinsic ceiling, not a training
gap, and it makes two consequences unavoidable. Absolute precision
will be low because the base rate of future coauthorship is tiny, and
a subset of realized coauthorships --- those whose topical alignment
has not yet reached the published record, or whose formation channel
is entirely outside it --- will never be retrievable by this test.
Our target use-case is therefore a \emph{discovery tool} for human
review, not an automation tool.

The ceiling does not make the test uninformative. If the semantic signal
carries real collaboration-relevant structure beyond what the coauthor
graph and publication volume already encode, then a predictor that
isolates that signal should beat popularity-weighted and same-venue
baselines by a substantive margin and should rank partners monotonically
by similarity. Both properties are directly testable. The remainder of
Sec.~\ref{sec:phantom} formalises the predictor
(Sec.~\ref{subsec:phantom-def}), runs the test
(Secs.~\ref{subsec:phantom-setup}--\ref{subsec:phantom-results}), and
illustrates its working envelope with a correct-case class and a
missed-pair class in Sec.~\ref{subsec:phantom-cases}. The reader should
therefore evaluate the phantom predictor on lift and calibration
gradient, not on absolute precision.

\subsection{Definition}
\label{subsec:phantom-def}

We now ask the predictive question formally. Given the state of the
field at some cutoff year $Y$, can a semantic-similarity score identify
pairs of authors who have \textbf{not yet} coauthored but will coauthor
in the future? We formalize this as a \textbf{phantom-collaborator} test.

Let $\mathbf{a}^{\leq Y}(a)$ denote the citation-weighted hybrid
centroid of author $a$ (the embedding $\mathbf{h}_i$ of
Sec.~\ref{subsec:semantic-centroids}, recomputed from only those
papers with publication year at most $Y$). Let $G^{\leq Y}$ denote
the train-period coauthor graph built from papers with year at most
$Y$. For each author $a$ we define the set of $K$ phantom partners as
\begin{equation}
  \mathcal{P}_K^{\leq Y}(a)
  = \bigl\{ b \in \mathrm{top}\text{-}K\, \cos(\mathbf{a}^{\leq Y}(a), \mathbf{a}^{\leq Y}(b))
    \,\big|\, d_{G^{\leq Y}}(a, b) \geq 3 \bigr\},
  \label{eq:phantom-def}
\end{equation}
where $\mathrm{top}\text{-}K$ returns the $K$ authors of highest cosine similarity to
$a$ excluding $a$ itself, and $d_{G}(a, b)$ is the geodesic distance in
$G$. The $\geq 3$ constraint excludes authors who are already coauthors
($d = 1$) or share a direct coauthor ($d = 2$), isolating pairs whose
future collaboration would not be predictable from the local social
structure alone.

A phantom pair $(a, b)$ is \emph{realized} in the test window
$[Y{+}1, 2025]$ if $a$ and $b$ coauthor at least one paper in that window.
The precision at $K$ of the phantom predictor is
\begin{equation}
  \mathrm{P}@K
  = \frac{\lvert\bigcup_a \{(a, b) : b \in \mathcal{P}_K^{\leq Y}(a),
                     \text{ realized}\}\rvert}
         {\sum_a \lvert \mathcal{P}_K^{\leq Y}(a) \rvert}.
  \label{eq:precision}
\end{equation}

\subsection{Temporal hold-out and baselines}
\label{subsec:phantom-setup}

We fix $Y = 2019$, leaving a six-year hold-out window from $2020$ to
$2025$. Crucially, \textbf{both} the hybrid embedding and the coauthor
graph are rebuilt from scratch on the pre-$2020$ subset. The SPECTER2
paper vectors are cached and therefore identical in both windows, but the
Arora whitening (Sec.~\ref{subsec:semantic-whitening}) is refitted on the
train subset only, so the top principal direction that is removed is the
\emph{pre-$2020$} common direction, not one that leaks information from
the hold-out. The train subset contains $\num{73586}$ papers with
embeddings and yields $\num{27713}$ authors eligible for evaluation
(at least two train-period papers and non-trivial centroid), of whom
$\num{5225}$ acquire at least one realized coauthor in the hold-out
window.

We compare the phantom predictor against three baselines, each sampling
$K$ candidate partners per anchor from a distinct null model, with all
near-train coauthors ($d \leq 2$) excluded.

\begin{itemize}
  \item \textbf{Random} uniformly samples from the pool of eligible
        authors. This is the weakest null and controls only for the
        base rate of future coauthorship.
  \item \textbf{Popularity-weighted} samples proportional to an author's
        train-period paper count. This is a pragmatic stand-in for a
        degree-preserving configuration-model null, which would produce
        the same first-order effect of popular authors attracting more
        new ties.
  \item \textbf{Same-venue} samples from the pool of authors who share
        at least one train-period publication venue with the anchor.
        This controls for opportunistic collaboration mediated by
        conference attendance or editorial networks.
\end{itemize}

\subsection{Results}
\label{subsec:phantom-results}

\begin{figure}[width=0.99\textwidth, pos = t]
  \centering
  \includegraphics[width=0.65\linewidth]{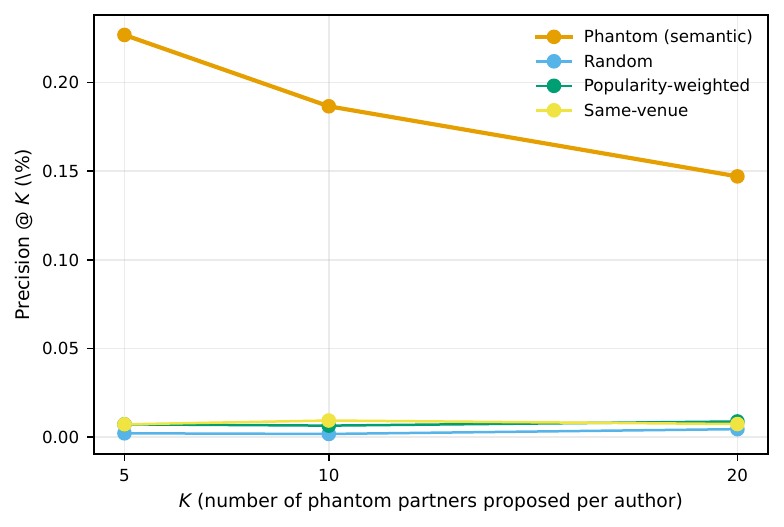}
  \caption{Precision at $K$ for the phantom predictor and three null
    baselines.}
  \label{fig:phantom-pk}
\end{figure}

\Cref{fig:phantom-pk} plots the precision at $K \in \{5, 10, 20\}$ for the
phantom predictor and the three baselines. The phantom curve dominates at
every $K$, and the gap widens at smaller $K$ where the top-ranked
phantoms are most discriminative.

\Cref{tab:phantom-lift} reports the exact counts, micro- and
macro-averaged precision, and the multiplicative lift of the phantom
predictor over each baseline. The two averages answer slightly
different questions. \textbf{Micro-precision} pools all predicted
pairs across all anchor authors and reports hits divided by total
predictions; it weights every prediction equally, so authors who
produce many predicted pairs contribute proportionally more. That is
the relevant figure for a deployed recommender that ranks pairs across
the whole user base. \textbf{Macro-precision} computes the hit rate
per anchor author first and then averages across authors; it weights
every author equally regardless of how many predicted pairs they
generated. That is the relevant figure if one cares about the typical
author's experience rather than the aggregate throughput. For the
phantom predictor the two averages agree within $0.01$ percentage
points because every anchor receives exactly $K$ predictions after the
$d \geq 3$ filter, so the per-author and aggregate distributions
coincide up to the small number of anchors with fewer than $K$
eligible candidates.

\begin{table}[width=0.99\textwidth, pos = t]
  \centering
  \small
  \caption{Phantom-predictor performance and lift over baselines.}
  \label{tab:phantom-lift}
    \begin{tabular}{lrrrrrr}
  \toprule
  \textbf{$K$} & \textbf{Method} & \textbf{Hits} & \textbf{Predictions} & \textbf{micro-P (\%)} & \textbf{macro-P (\%)} & \textbf{Lift vs phantom} \\
  \midrule
  5 & \textbf{Phantom (semantic)} & 311 & 137,160 & \textbf{0.23} & 0.23 & -- \\
   & Random & 3 & 138,565 & 0.00 & 0.00 & $\times$ 104.73 \\
   & Popularity-weighted & 10 & 138,565 & 0.01 & 0.01 & $\times$ 31.42 \\
   & Same-venue & 10 & 138,565 & 0.01 & 0.01 & $\times$ 31.42 \\
  \midrule
  10 & \textbf{Phantom (semantic)} & 501 & 268,591 & \textbf{0.19} & 0.20 & -- \\
   & Random & 5 & 277,130 & 0.00 & 0.00 & $\times$ 103.39 \\
   & Popularity-weighted & 18 & 277,130 & 0.01 & 0.01 & $\times$ 28.72 \\
   & Same-venue & 26 & 277,130 & 0.01 & 0.01 & $\times$ 19.88 \\
  \midrule
  20 & \textbf{Phantom (semantic)} & 638 & 433,988 & \textbf{0.15} & 0.19 & -- \\
   & Random & 25 & 554,260 & 0.00 & 0.00 & $\times$ 32.59 \\
   & Popularity-weighted & 49 & 554,260 & 0.01 & 0.01 & $\times$ 16.63 \\
   & Same-venue & 41 & 554,260 & 0.01 & 0.01 & $\times$ 19.87 \\
  \bottomrule
  \end{tabular}
\end{table}

At $K = 20$, the phantom predictor realizes at $\mathbf{0.1470\%}$, which
exceeds the random baseline by $32.6\times$, the same-venue baseline by
$19.9\times$, and the popularity-weighted baseline by $16.6\times$.
The popularity-weighted comparison is the most stringent: a
preferential-attachment draw already concentrates candidates on prolific
authors, which is the strongest null in the link-prediction literature
\citep{libennowell2007link, martinez2016survey}. Same-venue comparison is
the next most stringent, because authors who share a venue with the
anchor have already demonstrated editorial overlap and physical
proximity at conferences. Beating both by more than an order of
magnitude is the headline result of this section.

\begin{figure}[width=0.99\textwidth, pos = t]
  \centering
  \includegraphics[width=0.85\linewidth]{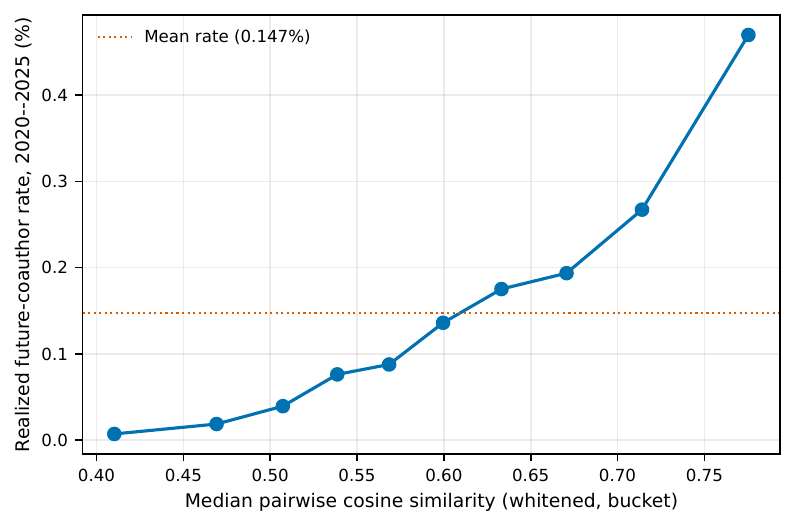}
  \caption{Relationship between pairwise cosine similarity and the
    realized future-coauthor rate at $K=20$.}
  \label{fig:phantom-calibration}
\end{figure}

Fig.~\ref{fig:phantom-calibration} shows that
the phantom score is an ordinal predictor rather than a binary threshold.
We split the top-$20$ semantic pairs into ten equal-frequency quantile
buckets of pairwise cosine similarity and compute the realized-rate in
each bucket. The top bucket realizes at $0.470\%$ and the bottom at
$0.0069\%$, a $\mathbf{68\times}$ monotone gradient. Higher similarity is
therefore associated with higher future-coauthorship probability across
the full range of the score.

\paragraph{Robustness to the $d \geq 3$ constraint.}
Equation~\eqref{eq:phantom-def} excludes both direct coauthors ($d = 1$)
and pairs sharing a coauthor ($d = 2$), isolating predictions that bridge
genuine social distance. Relaxing the constraint to $d \geq 2$ ---
that is, admitting triadic-closure cases (anchor and partner share a
coauthor in the train graph but have not directly coauthored) ---
increases the count of eligible anchors with at least one realized
phantom partner from $\num{5225}$ to $\num{6006}$ ($+14.9\%$) and
approximately doubles raw phantom precision: $0.45\%$ at $K = 5$,
$0.36\%$ at $K = 10$, and $0.28\%$ at $K = 20$, with $\num{1342}$
realized predictions at $K = 20$ versus $\num{638}$ under the strict
definition. Because the random and popularity-weighted baselines
remain near zero, the lift over random at $K = 5$ \emph{rises} from
$\times 105$ to $\times 206$, and the lift over popularity-weighted at
$K = 20$ rises from $\times 16.6$ to $\times 25.6$. The lift over the
same-venue baseline is essentially unchanged ($\times 19.9 \rightarrow
\times 26.0$ at $K = 20$), confirming that the predictor is not merely
re-discovering venue-mediated clustering. The headline finding is
therefore robust to the social-distance cutoff: when triadic-closure
cases are admitted, raw precision approximately doubles but the
multiplicative advantage of semantic similarity over null baselines is
preserved, and in several cells grows. We retain $d \geq 3$ as the
reported definition because it isolates the harder, more
discovery-relevant claim --- predicting collaborations that bridge social
distance rather than predicting closure of pre-existing acquaintance.

These two results together \textbf{answer RQ3} in the affirmative.
Semantic proximity predicts future coauthorship in a temporally
held-out evaluation, and the $16.6\times$ lift over popularity-weighted
and $19.9\times$ lift over same-venue baselines isolate a signal that
neither degree-based nor editorial-overlap nulls carry. The $68\times$
monotone calibration gradient shows that the score is ordinal rather
than a binary threshold, which is the property a discovery tool
requires.

\subsection{Case studies}
\label{subsec:phantom-cases}
\subsubsection{Correct Case}

The phantom-collaborator test produced $638$ realized pairs at $K = 20$
(\Cref{tab:phantom-lift}). \Cref{tab:phantom-cases} shows the highest-similarity slice.

Three named pairs illustrate the dominant success pattern. Row~$1$
(Lajunen--Stephens, $\cos = 0.905$, $d = 3$) links two established
driver-behavior researchers with $53$ and $31$ pre-$2020$ papers at
Finnish/Turkish and Australian institutions; they first coauthored in
$2025$. Row~$5$ (Winter--Krems, $\cos = 0.863$, $d = 3$) connects the TU
Delft and TU Chemnitz automated-driving groups ($30$ and $31$ pre-$2020$
papers), and Row~$6$ (Lee--Merat, $\cos = 0.862$, $d = 3$) is
transatlantic, pairing University of Wisconsin--Madison and University of
Leeds autonomous-driving human-factors research ($40$ and $25$ papers);
both Winter--Krems and Lee--Merat coauthored their first joint paper in
$2021$, within two years of the snapshot. In every case the predictor
surfaced a pair whose centroids had already converged and whose only
remaining gap was social.

Across the full top-$20$ realized slice, $11$ pairs sit at the minimum
eligible distance $d = 3$ and $9$ at $d \geq 4$, so the predictor
operates across the full eligibility range rather than only at the
boundary. Roughly half of the top-$20$ pairs involve driver-behavior or
road-safety researchers, a subtopic concentration that reflects a
coherent community with dense topical overlap but sparse cross-group
coauthorship --- exactly the setting a semantic recommender should help.
Seven of the $20$ pairs connect authors on different continents, which
is the positive counterpart of the geography-driven coauthor--semantic
decoupling discussed in Sec.~\ref{subsec:multiplex-cases}: the pairs
share topic but not institution, and the predictor closes that gap.

These successes share a structural pattern we call \emph{peer
convergence}: both authors in each pair have $\geq 25$ pre-$2020$
papers, so their centroids are stable and their topical agreement is
the genuine signal the predictor is designed to exploit, not an
artifact of a single outlier paper. Whatever historical mentorship ties
may exist among these pairs, by $2019$ both partners were already
independent established researchers. \Cref{tab:phantom-cases}
lists the top-$20$ deduplicated pairs, which is the full set a
reader needs to audit the predictor's working envelope.

\begin{table}[width=0.99\textwidth, pos = t]
  \centering
  \small
  \caption{Highest-similarity realized phantom pairs at $K=20$.}
  \label{tab:phantom-cases}
    \begin{tabular}{lll rr}
  \toprule
  \textbf{\#} & \textbf{Author A} & \textbf{Author B} & \textbf{Cos sim} & \textbf{Train dist} \\
  \midrule
  1 & Lajunen, Timo & Stephens, Amanda N. & 0.905 & 3 \\
  2 & Liu, Liu & Ai, Bo & 0.879 & $\geq 4$ \\
  3 & Zhou, Tao & Ai, Bo & 0.879 & $\geq 4$ \\
  4 & Miller, John M. & Mohamed, Ahmed A. S. & 0.870 & $\geq 4$ \\
  5 & Winter, Joost De & Krems, Josef F. & 0.863 & 3 \\
  6 & Lee, John D. & Merat, Natasha & 0.862 & 3 \\
  7 & Li, Shengbo Eben & Li, Yongfu & 0.857 & $\geq 4$ \\
  8 & Gong, Siyuan & Wang, Meng & 0.851 & 3 \\
  9 & Lee, John D. & Winter, Joost De & 0.851 & 3 \\
  10 & Horrey, William J. & Dingus, Thomas A. & 0.847 & 3 \\
  11 & Moller, Mette & Stephens, Amanda N. & 0.846 & $\geq 4$ \\
  12 & Li, Changle & Cheng, Nan & 0.844 & 3 \\
  13 & Grzebieta, Raphael & Haworth, Narelle & 0.843 & 3 \\
  14 & Martens, Marieke & Krems, Josef F. & 0.842 & 3 \\
  15 & Ben-Ari, Orit Taubman - & Ozkan, Turker & 0.838 & 3 \\
  16 & Hui, Yilong & Cheng, Nan & 0.838 & 3 \\
  17 & Cheng, Nan & Su, Zhou & 0.838 & 3 \\
  18 & Lee, Loo Hay & Lee, Chung-Yee & 0.835 & 3 \\
  19 & He, Zhengbing & Li, Li & 0.834 & 3 \\
  20 & Scott-Parker, Bridie & Simons-Morton, Bruce G. & 0.833 & $\geq 4$ \\
  \bottomrule
  \end{tabular}
\end{table}

\subsubsection{Missed Case}
An exemplar missed pair tells a different story: the anchor is still
career-thin, and the anchor's latent research trajectory has not yet
surfaced in their published record because publishing lags formation of
interest by several years. The predictor --- which sees only what has
been published --- therefore cannot see the topical fit that the anchor
is already pursuing. Missed pairs are also where the upper bound of
what a top-$20$ retrieval cutoff can return becomes visible, and a
transparent missed-pair discussion is more persuasive than a
curated-hits table. We illustrate the ceiling with our own case:
the first author of this paper (S.~Choi) and L.~Sun (the first author of
\citet{sun2017coauthorship}), whose first coauthored paper was published
in $2024$ after S.~Choi joined L.~Sun's lab as a postdoctoral researcher
in $2022$.

In the pre-$2020$ train graph the two sat $5$ hops apart, clearing the
$d \geq 3$ phantom eligibility filter. Their pairwise cosine in the
$2019$ hybrid space was $0.205$, which placed L.~Sun at rank
$\mathbf{4{,}147}$ of $\num{27712}$ eligible neighbors of S.~Choi. The
similarity needed to enter Choi's top-$20$ in $2019$ was $0.599$, and
the $0.394$ gap between that cutoff and L.~Sun's actual similarity is substantial rather than marginal. Eq.~\ref{eq:phantom-def} therefore does not retain the pair,
and it falls outside the $2020$--$2025$ realized-pair hold-out on which
the precision numbers above are measured.

Two independent forces produced the miss, and only one of them is a
predictor-side issue. S.~Choi was a doctoral student in $2019$ with only
$2$ train-period papers on connected-vehicle control, so his centroid
was high-variance and centered on a narrow subtopic: three of his top
five pre-$2020$ semantic neighbors were already direct coauthors in the
train graph ($d=1$), reflecting a publication base anchored in a single
research group rather than a diversified topical span. L.~Sun's centroid,
by contrast, was built from $16$ train-period papers already anchored in
urban-mobility data fusion. Their topical overlap in $2019$ was genuinely
small; Choi's shift toward urban-mobility data and AI for transportation
was \emph{under way} in $2019$ but had not yet surfaced in his
published record because publication lags research-interest formation
by several years. This is the predictor-side force. The second force
is temporal: a text-based predictor can only see \emph{already
published} work, so it cannot see a researcher who is actively seeking
a topical fit that their existing publications do not yet represent.
The collaboration formed because Choi identified the fit himself and
joined Sun's lab to pursue it; by the time that pursuit surfaced as
coauthored papers, the $2019$-snapshot predictor was years out of
date.

Naming these forces defines the ceiling of the phantom test. The
predictor systematically under-serves two classes of anchor: doctoral
researchers with thin publication records and high-variance centroids,
and researchers whose nascent research interests have not yet surfaced
in their published work. Row $1$ of
Table~\ref{tab:phantom-cases} (Lajunen--Stephens, $\cos = 0.905$,
$d = 3$) is the opposite case: two established authors whose stable
centroids and dense publication records make a high-confidence prediction
possible. Cases of the Choi--Sun form are part of why the low absolute precision of Sec.~\ref{subsec:phantom-results} is expected rather than pathological, and they motivate the rolling-cutoff evaluation
in Sec.~\ref{subsec:disc-future}: recomputed annually, Choi's centroid
would drift toward Sun's as post-$2019$ urban-mobility papers enter the
corpus, and a rolling phantom predictor would be positioned to surface
such pairs before the human pathway closes them.

\subsubsection{Implications}
These results have three implications. First, the hybrid embedding
captures substantively more collaboration-relevant structure than either
coauthorship or shared venue alone. Second, the low absolute precision
($0.15$--$0.23\%$) means phantom prediction is a discovery tool rather
than an automation tool. Our target use-case is an interactive
recommender inside the atlas that surfaces candidate partners, not a
system that fires off introductions autonomously. Third, the predictive
signal justifies treating the atlas as a live research artifact. Because
new papers continually update author centroids, the set of phantoms is
refreshed automatically whenever the pipeline runs.

\section{Author topic trajectories}
\label{sec:trajectories}

This section presents a career-arc view of the corpus that
complements the pair-level phantom-collaborator test of
Sec.~\ref{sec:phantom}. The phantom test asks ``which two authors
\emph{should} write a paper together'' by comparing centroids at a
single snapshot. The trajectory taxonomy below asks the dual
career-level question, ``how does an author's topical position evolve
over decades?'', by tracking each author's centroid across five-year
bins and classifying the resulting polyline by shape. The
construction is also what backs the interactive five-year
trajectory-playback view in the atlas (Sec.~\ref{sec:atlas}).

\subsection{Trajectory construction}
\label{subsec:traj-construction}

An author's topic can change over time. To quantify this we partition
each author's papers into five-year bins anchored at their first
observed publication year and compute a bin-level citation-weighted
centroid in the UMAP-projected space of
Sec.~\ref{subsec:semantic-centroids}. We keep authors who have at
least two non-empty bins, leaving $\mathbf{8{,}589}$ trajectories
(we drop the partial 2025--2029 bin because it is only partly
populated at the snapshot date). Each trajectory is a polyline in the
$2$-D UMAP plane whose segments carry the author through adjacent
career phases.

For each trajectory we compute two scalars. The \textbf{total path}
is the sum of consecutive segment lengths $\sum_{i} \|\mathbf{p}_{i+1}
- \mathbf{p}_{i}\|$, and measures the total distance traveled
regardless of direction. The \textbf{net displacement} is the
end-to-end distance $\|\mathbf{p}_{\text{last}} -
\mathbf{p}_{\text{first}}\|$. Their ratio
$\eta = \mathrm{net} / \mathrm{path}$ lies in $[0, 1]$ and equals one
when the trajectory is a straight line and approaches zero when the
author repeatedly returns near earlier positions.

\subsection{Stayer / drifter / returner / switcher taxonomy}
\label{subsec:traj-taxonomy}

Total path and efficiency cleanly separate the bulk profiles, but
inside the low-$\eta$ tail the \emph{shape} of the path matters: a
trajectory that wanders out and comes home tells a different story
from one that re-orients to a new home permanently. Net
displacement~$d = \lVert \mathbf{c}_{T} - \mathbf{c}_{1} \rVert$
disambiguates the two. We therefore assign each author to one of
\emph{four} classes.

\begin{itemize}
  \item \textbf{Stayers} have total path below $15$ UMAP units. Their
        research center of mass barely moves between bins, either
        because they have short careers or because they remain in the
        same subfield throughout.
  \item \textbf{Drifters} have total path above $15$ UMAP units and
        $\eta \geq 0.60$. Their trajectory is long but roughly
        monotonic, consistent with a gradual migration toward an
        adjacent subfield.
  \item \textbf{Returners} (round-trip pivoters) have total path
        above $15$, $\eta < 0.60$, and \emph{net displacement}
        $d < 15$. Their trajectory is long and non-monotonic, but the
        path closes back near its origin --- the canonical
        ``excursion and return'' shape.
  \item \textbf{Switchers} (re-oriented authors) have total path
        above $15$, $\eta < 0.60$, and net displacement $d \geq 15$.
        Their trajectory is also non-monotonic, but the author ends
        up in a different semantic neighborhood than where they
        started --- a mid-career relocation rather than a round-trip.
\end{itemize}

The two pivoter sub-classes share the same $\eta$ floor but differ in
where the path lands; empirically the $\eta$ distribution is sharply
bimodal (pivoter $q_{75} = 0.42$, drifter $q_{05} = 0.68$, with
almost no mass in between), so the four-class partition is robust to
small changes in the cutoffs.

\begin{figure}[width=0.99\textwidth, pos = h]
  \centering
  \includegraphics[width=0.75\linewidth]{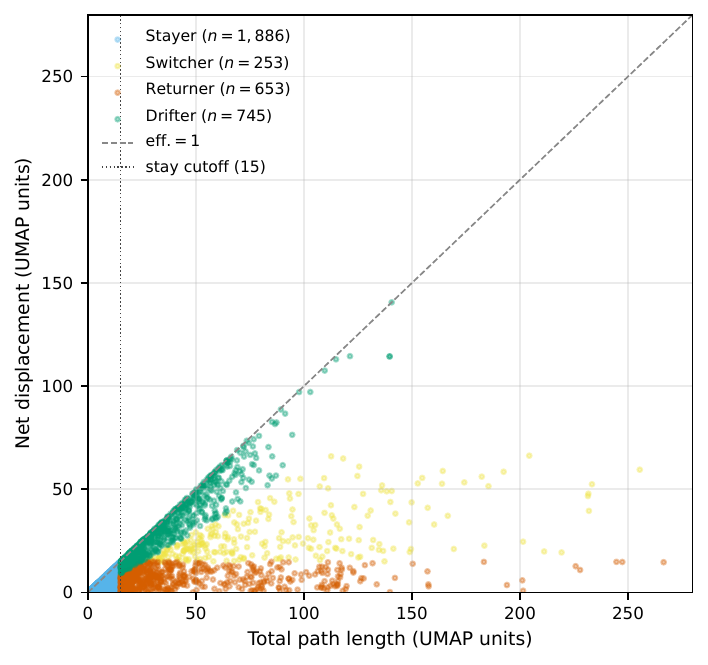}
  \caption{Total path length versus net displacement (UMAP units),
    restricted to the $\num{3537}$ authors with $\geq 3$ five-year
    bins.}
  \label{fig:traj-scatter}
\end{figure}

\paragraph{Restriction to shape-rich trajectories.}
Of the $\num{168090}$ unique authors in the corpus, $\num{41551}$
have at least two papers and enter the coauthor graph
(Sec.~\ref{sec:coauthor}). The trajectory analysis further requires
that the author's papers span at least two 5-year bins so that a
centroid path is defined; this leaves $\num{8589}$
trajectory-eligible authors. Of those, $5{,}052$ ($59\%$) have only
\emph{two} bins, and 2-bin trajectories are degenerate: the path is
a single segment, so $\eta = 1$ by construction and the returner and
switcher classes are structurally unreachable. We therefore restrict
the headline analysis to authors with at least \emph{three} bins ---
a $\geq\!10$-year active publication window in our corpus, the
minimum bin count for which the path can have $\geq 2$ segments and
thus register a non-monotone shape. The shape-rich subset contains
$\mathbf{3{,}537}$ authors, partitioned into $\mathbf{53.3\%}$
stayers, $\mathbf{21.1\%}$ drifters, $\mathbf{18.5\%}$ returners, and
$\mathbf{7.2\%}$ switchers; the returner and switcher counts are
identical to those in the unrestricted set by construction, since
a 2-bin trajectory has $\eta = 1$ and so cannot fall into either class.
\Cref{fig:traj-scatter} plots total path against net displacement for
every shape-rich author, colored by class, with the $\eta = 1$
diagonal and the $\mathrm{path} = 15$ vertical cutoff shown for
reference. \Cref{tab:traj-stats} reports per-class medians of the
path, net, efficiency, and several career covariates over the same
subset.

\begin{table}[width=0.99\textwidth, pos = h]
  \centering
  \small
  \caption{Per-class trajectory statistics (medians) over the
    shape-rich subset.}
  \label{tab:traj-stats}
    \begin{tabular}{lrrrrrrrr}
  \toprule
  \textbf{Class} & \textbf{$n$} & \textbf{path} & \textbf{net} & \textbf{eff.} & \textbf{papers} & \textbf{cites} & \textbf{span (yr)} & \textbf{bins} \\
  \midrule
  Stayer    & 1{,}886 &  6.2 &  3.3 & 0.67 & 16 & 557 & 15 & 3 \\
  Drifter   &    745  & 32.1 & 26.9 & 0.89 & 17 & 530 & 20 & 3 \\
  Returner  &    653  & 28.9 &  6.9 & 0.19 & 21 & 642 & 25 & 4 \\
  Switcher  &    253  & 72.2 & 23.7 & 0.38 & 21 & 620 & 25 & 4 \\
  \bottomrule
  \end{tabular}
\end{table}

\subsection{Correlates}
\label{subsec:traj-correlates}

Stayers in the shape-rich subset are no longer the one-bin-per-paper
short-career authors who dominated the 2-bin tail; the median stayer
here has $16$ papers across a $15$-year publication window, but the
citation-weighted centroid never moves more than $\tau_{\mathrm{stay}}
= 15$ UMAP units. Drifters publish at a similar rate ($17$ papers,
$20$-year span) but their trajectory traverses substantially more
ground in roughly a straight line. Returners and switchers are the
most prolific classes ($21$ papers each, $25$-year careers) and have
comparable citation totals; they differ from each other not in
productivity but in trajectory shape --- returners come back to
their starting neighborhood, switchers do not.

\begin{figure}[width=0.99\textwidth, pos = t]
  \centering
  \includegraphics[width=\linewidth]{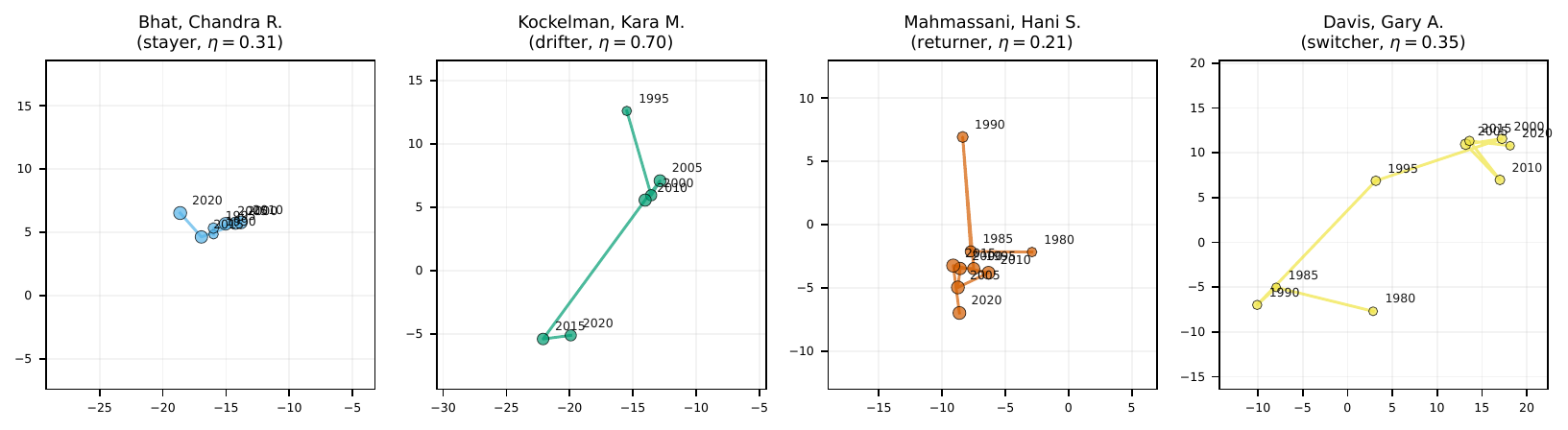}
  \caption{Exemplar trajectories.}
  \label{fig:traj-examples}
\end{figure}

\Cref{fig:traj-examples} grounds the taxonomy in four household-name
transportation researchers, one per class. Each panel plots the
UMAP-projected centroid for one author across their five-year career
bins; panel titles give the author's class label and computed
$\eta$. The four trajectories make the formal cutoffs interpretable
in domain terms.

\paragraph{Stayer.} \textit{Bhat, Chandra~R.} is classified as a stayer.
His seven bins from
$1995$ through $2020$ all sit inside the discrete-choice and
activity-based travel-demand neighborhood (stated-preference
modeling, car-use behavior, activity scheduling), and each bin's
most-cited papers continue to develop the same methodological
program he established in the mid-$1990$s. Later work extends into
MaaS adoption and well-being but stays within the broader behavioral
travel-demand family. 

\paragraph{Drifter.} \textit{Kockelman, Kara~M.} is classified as a drifter.
Her early bins ($1995$--$2010$) sit alongside Bhat in
land-use and travel-demand modeling territory, while her later bins
($2015$, $2020$) have migrated to connected- and automated-vehicle
adoption, electric-vehicle fleet operations, and shared-mobility
policy. The migration is monotonic: each successive five-year
centroid is closer to the eventual CAV-adoption neighborhood than
the previous one, with no reversal back toward travel-demand
modeling. 

\paragraph{Returner.} \textit{Mahmassani, Hani~S.} is classified as a returner.
His career anchor is dynamic traffic assignment and
network-level traffic flow modeling and his $1985$, $1995$, and every
subsequent bin cluster tightly inside this neighborhood. His $1990$
bin is the only one that does not: it jumps cleanly into the
early-$1990$s research wave on advanced traveler information systems
(ATIS) and intelligent vehicle-highway systems (IVHS) --- real-time
information provision, dynamic shortest-path algorithms for IVHS, and
commuter response to information (route switching, telecommuting
adoption) --- far enough that this single segment contributes more to
total path than any other in his career, and by $1995$ he is back in
dynamic-assignment territory for the next thirty years. 


\paragraph{Switcher.} \textit{Davis, Gary~A.} is classified as a switcher.
His $1980$s bins lie in freeway-traffic operations and flow analysis;
his mid-$1990$s bin marks a sharp shift into a different research
family --- statistical methods for highway-safety analysis, crash
modeling, and safety-performance functions for rural highways and
urban arterials --- and his $2000$--$2020$ bins all settle into this
safety neighborhood and stay there. The total path ($67.8$ units,
almost twice Mahmassani's) is large, and unlike Mahmassani the
trajectory does not close: net displacement is $24$ units, and
$\eta = 0.35$ is low not because the path returns home but because
the early operations-research direction and the later
safety-research direction partially cancel along orthogonal axes of
the embedding. 

These patterns broadly agree with the folklore that strong
researchers either double down on a narrow question or deliberately
change gears every decade or so. The taxonomy formalises the
folklore. It also gives the atlas a handle for exploration by career
arc rather than by current position, which complements the
phantom-based recommendations of Sec.~\ref{sec:phantom}.

\section{The Transport Atlas tool}
\label{sec:atlas}

The analyses in Secs.~\ref{sec:descriptive}--\ref{sec:phantom} are
backed by a public interactive atlas at
\url{https://choi-seongjin.github.io/transport-atlas/}. The atlas is a
static site with six views. \textbf{Explorer} offers a Tabulator.js
table of all papers with free-text search, per-venue filtering, and
citation-count sorting. \textbf{Papers by Year} overlays the stacked
annual counts used in Fig.~\ref{fig:papers-by-year} with per-venue
toggles and a Chart.js breakdown by author. \textbf{Coauthor Network}
renders the Leiden-colored graph of
Sec.~\ref{sec:coauthor} with Sigma.js and Graphology, supporting
pan, zoom, and click-to-highlight navigation. \textbf{Topic Space}
renders the UMAP of Sec.~\ref{sec:semantic} with the same interactions
plus $5$-year trajectory playback. \textbf{Trajectories} surfaces the
four-class career-arc taxonomy of Sec.~\ref{sec:trajectories} as an
interactive total-path versus net-displacement scatter; clicking a
point or a class-filtered list row reveals that author's
year-labeled UMAP polyline alongside their class label and computed
$\eta$. A \textbf{Combined} view overlays the coauthor force layout
with semantic-community coloring and adds phantom-collaborator pairs
(Sec.~\ref{sec:phantom}) as amber edges, surfacing semantically
close authors who have not coauthored.

The pipeline is implemented in Python 3.11. Storage uses Parquet read
through DuckDB. Graph construction runs in NetworkX, community detection
runs in \texttt{python-igraph} with \texttt{leidenalg}, UMAP uses
\texttt{umap-learn}, and paper embeddings are computed with the
\texttt{allenai/specter2} model in PyTorch on a single GPU. The
frontend is CDN-hosted Chart.js, Tabulator, and Sigma.js without any
build step, so the site works directly from GitHub Pages. All metadata,
embeddings, and derived community assignments ship under the OpenAlex
CC0 and Crossref CC0 licenses. Source code is released at
\url{https://github.com/UMN-Choi-Lab/transport-atlas}. The snapshot
corresponding to the figures in this paper was taken on
\textit{2026-04-23}.

\section{Discussion}
\label{sec:discussion}

\subsection{Limitations}
\label{subsec:disc-limits}

Certain limitations should be noted. First, our corpus is English-language
only, because SPECTER2 is trained predominantly on English scientific text
and because OpenAlex coverage of non-English venues is incomplete in the
decades before $2015$. Chinese, Japanese, and Korean transportation
journals are therefore under-represented, both in the descriptive tables
of Sec.~\ref{sec:descriptive} and in the semantic structure of
Sec.~\ref{sec:semantic}. Second, OpenAlex coverage of pre-$2000$ papers
varies sharply across venues, with some venues achieving $40$--$80\%$
rather than near-full coverage. We quantify the per-venue coverage in
Appendix~\ref{app:coverage}. Where coverage is thin we necessarily
under-count collaborations and productivity. Third, author
disambiguation is imperfect. The chain we use
(OpenAlex ID, ORCID, canonical key) is robust on average but produces
residual errors on common Chinese and Korean surnames. During development
we caught and corrected a five-way split involving the surname ``Sun,
Lijun'' by adding a manual alias to the pipeline, and similar cases
probably remain in the long tail. Fourth, SPECTER2 is a frozen model.
Re-running the pipeline with a different embedding model will shift the
semantic communities and the phantom set. We pin the model version and
ship the whitening PC so that results are reproducible. Fifth, the
phantom recommender risks amplifying existing biases. Authors who
currently receive disproportionate visibility will tend to appear as
phantoms of each other. Before any deployment we therefore recommend an
equity-of-exposure audit in the spirit of fair ranking.

\subsection{Policy implications}
\label{subsec:disc-policy}

The atlas and its derived measures suggest several actionable uses.
\textbf{Panel composition} for conference tracks or funding review can
aim for diversity across coauthor communities (Sec.~\ref{sec:coauthor})
and across semantic communities (Sec.~\ref{sec:semantic}) separately.
Relying on either dimension alone loses the other. \textbf{Cross-community
collaboration metrics} for funding agencies can use the bridge-edge
count defined in Sec.~\ref{subsec:coauthor-bridges}. Programs that
reward brokerage of previously disjoint communities have an objective
yardstick. \textbf{Literature-review tooling} for early-career
researchers can use the trajectory taxonomy of
Sec.~\ref{sec:trajectories} to identify exemplars whose careers trace a
shape they may wish to emulate, independently of whose papers they
already know. \textbf{Candidate-partner discovery} can use the phantom
predictor in Sec.~\ref{sec:phantom} as a browsing aid, not as an
automated introduction system, given the absolute precision the
predictor currently achieves. All four applications are compatible with
the existing atlas front-end.

\subsection{Future work}
\label{subsec:disc-future}

Several natural extensions are under way. First, the current partitions
are static snapshots, and a time-aware dynamic-Leiden analysis
\citep{mucha2010community} would capture the emergence and dissolution
of communities that our giant-component plot only summarizes. Second,
the phantom test uses a single train/test split at $2019$. A rolling
evaluation across multiple cutoff years would quantify how far into the
future the predictor remains accurate and whether its precision
degrades systematically for pairs separated by more than five years of
non-overlap. Third, the same author centroids and coauthor adjacency
naturally support an editorial reviewer-matching tool. Fourth, the
fraction of each author's coauthors in their own semantic-Leiden
community --- a \textbf{collaboration-style} score, benchmarked against
a degree-preserving rewire null --- would test whether the field favors
topical homophily or heterophily, and whether the answer correlates
with citation impact or the career-arc shapes of
Sec.~\ref{sec:trajectories}.


\section{Conclusion}
\label{sec:conclusion}

We have presented a semantic-structural atlas of transportation research
covering $\num{120323}$ papers across $34$ venues from $1967$ to $2025$,
roughly an order of magnitude larger than the previous reference
bibliometric study \citep{sun2017coauthorship}. Beyond extending the
descriptive scaffold and the coauthor-network analysis, we add three
methodological layers that the atlas uniquely enables. We build a hybrid
author embedding combining whitened SPECTER2, OpenAlex concept TF-IDF,
and venue LDA, and use it to detect $23$ semantic communities that are
only weakly aligned with the coauthor partition. We combine the two into a multiplex graph
whose $181$ communities recover both layers with partial information
loss. Finally, we formalize a \textbf{phantom-collaborator} predictive
test
that identifies, for each author active by the hold-out year, the set of
semantically close but socially distant authors most likely to become
future coauthors.

The three research questions of Sec.~\ref{sec:intro} are answered as
follows. The field has grown by roughly an order of magnitude since the
\citet{sun2017coauthorship} window and has crossed the connectivity
threshold that distinguishes mature scientific communities. Topic and
collaboration structure agree only weakly, with normalized mutual
information of $0.23$ between the coauthor and semantic partitions,
opening room for the predictive layer. Semantic proximity is indeed a
calibrated predictor of future coauthorship. Phantom pairs realize
$16$ to $33$ times more often than three distinct null baselines, and
the realized-rate varies by a $68$-fold monotone gradient across
similarity buckets.


\section*{Data availability}
All paper metadata, author embeddings, and derived community
assignments are released under the OpenAlex CC0 and Crossref CC0
licenses at \url{https://github.com/UMN-Choi-Lab/transport-atlas}.

\section*{Declaration of competing interest}
The author declares that he has no known competing financial
interests or personal relationships that could have appeared to
influence the work reported in this paper.

\section*{Declaration of generative AI and AI-assisted technologies in the writing process}
During the preparation of this work the author used Anthropic Claude
(Sonnet and Opus models) and OpenAI ChatGPT to assist with code
refactoring, prose copy-editing, and proofreading of LaTeX source.
After using these tools the author reviewed and edited the content as
needed and takes full responsibility for the content of the
publication.

\bibliographystyle{cas-model2-names}
\bibliography{refs}

\appendix
\clearpage

\section{Venue ISSNs and coverage}
\label{app:venues}

\begin{table}[width=0.99\textwidth, pos = h]
  \centering
  \footnotesize
  \caption{The 34 venues indexed by the atlas. Coverage is the observed
    year span in OpenAlex for papers of type \texttt{journal-article} or
    \texttt{conference-paper}.}
  \label{tab:app-venues}
  \resizebox{\textwidth}{!}{
    \begin{tabular}{lllrl}
  \toprule
  \textbf{Venue} & \textbf{Abbr.} & \textbf{ISSN (print / online)} & \textbf{Papers} & \textbf{Coverage} \\
  \midrule
  Transportation Research Part A: Policy and Practice & TR-A & 0965-8564 / 1879-2375 & 5,013 & 1979--2025 \\
  Transportation Research Part B: Methodological & TR-B & 0191-2615 / 1879-2367 & 3,778 & 1979--2025 \\
  Transportation Research Part C: Emerging Technologies & TR-C & 0968-090X / 1879-2359 & 4,754 & 1993--2025 \\
  Transportation Research Part D: Transport and Environment & TR-D & 1361-9209 / 1879-2340 & 4,523 & 1996--2025 \\
  Transportation Research Part E: Logistics and Transportation Review & TR-E & 1366-5545 / 1878-5794 & 3,832 & 1997--2025 \\
  Transportation Research Part F: Traffic Psychology and Behaviour & TR-F & 1369-8478 / 1873-5517 & 3,464 & 1998--2025 \\
  IEEE Transactions on Intelligent Transportation Systems & T-ITS & 1524-9050 / 1558-0016 & 11,007 & 1997--2025 \\
  IEEE Transactions on Intelligent Vehicles & T-IV & 2379-8858 / 2379-8904 & 2,146 & 2016--2025 \\
  IEEE Intelligent Transportation Systems Magazine & ITS Mag & 1939-1390 / 1941-1197 & 1,457 & 2009--2025 \\
  Communications in Transportation Research & CommTR & 2772-4247 & 182 & 2021--2025 \\
  Transportation Science & TransSci & 0041-1655 & 2,552 & 1967--2025 \\
  Transportation Research Record: Journal of the Transportation Research Board & TRR & 0361-1981 / 2169-4052 & 38,423 & 1967--2025 \\
  Analytic Methods in Accident Research & AMAR & 2213-6657 / 2213-6665 & 277 & 2013--2025 \\
  Transport Reviews & TranspRev & 0144-1647 / 1464-5327 & 1,756 & 1981--2025 \\
  Journal of Transport Geography & JTG & 0966-6923 / 1873-1236 & 4,115 & 1993--2025 \\
  Accident Analysis and Prevention & AAP & 0001-4575 / 1879-2057 & 9,204 & 1969--2025 \\
  Journal of Urban Mobility & JUM & 2667-0917 & 167 & 2021--2025 \\
  Travel Behaviour and Society & TBS & 2214-367X / 2214-3688 & 1,117 & 2014--2025 \\
  Transport Policy & TranspPol & 0967-070X / 1879-310X & 3,578 & 1993--2025 \\
  Research in Transportation Business and Management & RTBM & 2210-5395 / 2210-5409 & 1,239 & 2011--2025 \\
  European Transport Research Review & ETRR & 1867-0717 / 1866-8887 & 744 & 2008--2025 \\
  International Journal of Sustainable Transportation & IJST & 1556-8318 / 1556-8334 & 989 & 2007--2025 \\
  Transportation Research Interdisciplinary Perspectives & TRIP & 2590-1982 & 1,642 & 2019--2025 \\
  Journal of Public Transportation & JPT & 1077-291X / 2375-0901 & 665 & 1996--2025 \\
  Computer-Aided Civil and Infrastructure Engineering & CACIE & 1093-9687 / 1467-8667 & 2,822 & 1986--2025 \\
  Journal of Intelligent and Connected Vehicles & JICV & 2399-9802 & 137 & 2018--2025 \\
  Journal of Traffic and Transportation Engineering (English Edition) & JTTE & 2095-7564 & 690 & 2014--2025 \\
  Railway Engineering Science & RailEng & 2662-4745 / 2662-4753 & 204 & 2020--2025 \\
  IEEE Open Journal of Intelligent Transportation Systems & OJ-ITS & 2687-7813 & 368 & 2019--2025 \\
  International Journal of Transportation Science and Technology & IJTST & 2046-0430 / 2046-0449 & 666 & 2012--2025 \\
  eTransportation & eTransp & 2590-1168 & 460 & 2019--2025 \\
  Green Energy and Intelligent Transportation & GEIT & 2097-2512 / 2773-1537 & 223 & 2022--2025 \\
  IEEE Transactions on Transportation Electrification & T-TE & 2332-7782 & 3,748 & 2015--2025 \\
  IEEE Vehicular Technology Magazine & VT Mag & 1556-6072 / 1556-6080 & 1,588 & 2006--2025 \\
  \bottomrule
  \end{tabular}
  }
\end{table}

\clearpage

\section{OpenAlex coverage by decade $\times$ venue}
\label{app:coverage}

\Cref{tab:app-coverage} reports paper counts by decade for all $34$
venues, sorted by total volume. Coverage is sparsest in the
$1960$s--$1980$s, when several venues had not yet launched, and
densest from the $2010$s onward across every venue family.

\begin{table}[width=0.99\textwidth, pos = h]
  \centering
  \footnotesize
  \caption{Paper counts by decade for all $34$ venues, sorted by total
    papers. The $1960$s column reflects the founding of
    \textit{Transportation Research Record} in $1967$.}
  \label{tab:app-coverage}
    \begin{tabular}{lrrrrrrr}
  \toprule
  \textbf{Venue} & \textbf{1960s} & \textbf{1970s} & \textbf{1980s} & \textbf{1990s} & \textbf{2000s} & \textbf{2010s} & \textbf{2020s} \\
  \midrule
  TR-A & -- & 1 & 18 & 738 & 604 & 1,914 & 1,738 \\
  TR-B & -- & 35 & 442 & 394 & 578 & 1,509 & 820 \\
  TR-C & -- & -- & -- & 180 & 350 & 2,030 & 2,194 \\
  TR-D & -- & -- & -- & 94 & 436 & 1,660 & 2,333 \\
  TR-E & -- & -- & -- & 73 & 446 & 1,287 & 2,026 \\
  TR-F & -- & -- & -- & 31 & 331 & 1,481 & 1,621 \\
  T-ITS & -- & -- & -- & 9 & 541 & 3,074 & 7,383 \\
  T-IV & -- & -- & -- & -- & -- & 261 & 1,885 \\
  ITS Mag & -- & -- & -- & -- & 62 & 747 & 648 \\
  CommTR & -- & -- & -- & -- & -- & -- & 182 \\
  TransSci & 88 & 342 & 310 & 348 & 386 & 614 & 464 \\
  TRR & 12 & 2,433 & 5,045 & 7,092 & 9,211 & 9,324 & 5,306 \\
  AMAR & -- & -- & -- & -- & -- & 104 & 173 \\
  TranspRev & -- & -- & 260 & 292 & 416 & 487 & 301 \\
  JTG & -- & -- & -- & 384 & 510 & 1,715 & 1,506 \\
  AAP & 134 & 490 & 716 & 914 & 1,486 & 3,386 & 2,078 \\
  JUM & -- & -- & -- & -- & -- & -- & 167 \\
  TBS & -- & -- & -- & -- & -- & 232 & 885 \\
  TranspPol & -- & -- & -- & 171 & 418 & 1,334 & 1,655 \\
  RTBM & -- & -- & -- & -- & -- & 406 & 833 \\
  ETRR & -- & -- & -- & -- & 20 & 357 & 367 \\
  IJST & -- & -- & -- & -- & 62 & 508 & 419 \\
  TRIP & -- & -- & -- & -- & -- & 70 & 1,572 \\
  JPT & -- & -- & -- & 43 & 226 & 249 & 147 \\
  CACIE & -- & -- & 127 & 371 & 467 & 672 & 1,185 \\
  JICV & -- & -- & -- & -- & -- & 19 & 118 \\
  JTTE & -- & -- & -- & -- & -- & 335 & 355 \\
  RailEng & -- & -- & -- & -- & -- & -- & 204 \\
  OJ-ITS & -- & -- & -- & -- & -- & 1 & 367 \\
  IJTST & -- & -- & -- & -- & -- & 226 & 440 \\
  eTransp & -- & -- & -- & -- & -- & 24 & 436 \\
  GEIT & -- & -- & -- & -- & -- & -- & 223 \\
  T-TE & -- & -- & -- & -- & -- & 393 & 3,355 \\
  VT Mag & -- & -- & -- & -- & 253 & 777 & 558 \\
  \midrule
  \textbf{Total} & \textbf{234} & \textbf{3,301} & \textbf{6,918} & \textbf{11,134} & \textbf{16,803} & \textbf{35,196} & \textbf{43,944} \\
  \bottomrule
  \end{tabular}
\end{table}

\clearpage
\section{Models, hyperparameters, and seeds}
\label{app:hyperparams}

This appendix consolidates every tunable that enters the pipeline from
ingest to phantom evaluation. The authoritative values live in
\texttt{config/pipeline.yaml}; the summary here exists so that a reader
who wants to reproduce one stage need not cross-reference the YAML.

\subsection{Ingest and deduplication}
\label{subsec:app-c-ingest}

\paragraph{OpenAlex.} Paginated via cursor with \texttt{per\_page}${=}200$
and a polite rate cap of $8$ requests per second. Each venue is resolved
by ISSN to an OpenAlex source ID, then checkpointed per-venue so that an
interrupted run resumes without re-downloading.

\paragraph{IEEE Xplore.} Used only to backfill abstracts and author
keywords for T-ITS, T-IV, ITS~Mag, and OJ-ITS before roughly $2010$,
where OpenAlex abstract coverage is sparse. Rate cap $5$~rps,
$200$ records per call, hard ceiling of $10{,}000$ records per venue.

\paragraph{Crossref.} DOI-level fallback only, hit through the polite
pool with a contact email.

\paragraph{Deduplication.} Pass~1 collapses rows sharing a normalized
DOI. Pass~2 applies \texttt{rapidfuzz} token-set ratio on the
lower-cased, punctuation-stripped title with threshold~$95$, and further
requires identical publication year and at least one overlapping author
surname (\texttt{min\_surname\_overlap = 1}). Both thresholds are
enforced jointly so that a fuzzy title alone cannot merge two papers.

\paragraph{Citation reconciliation.} Per paper,
\[
  \mathrm{citations}
  = \max\!\bigl(\texttt{openalex.cited\_by\_count},\,
                \texttt{crossref.is-referenced-by-count}\bigr),
\]
with missing values treated as zero before the max.

\subsection{Paper embeddings}
\label{subsec:app-c-embeddings}

\paragraph{SPECTER2.} \texttt{allenai/specter2}, base variant. Input is
the concatenated title and abstract with the model's default
\texttt{[SEP]} token separator. Papers missing an abstract fall back to
title-only input, which SPECTER2 tolerates at reduced fidelity.

\paragraph{Whitening.} The top-$1$ principal component, which alone
explains $9.6\%$ of the raw SPECTER2 corpus variance, is projected out
of the mean-centered embeddings, followed by a per-dimension $z$-score.
The median pairwise cosine similarity between two randomly drawn
corpus papers shifts from $0.833$ (raw) to $-0.010$ (whitened) on a sample of
$\num{199956}$ pairs drawn from $5{,}000$ papers
(Fig.~\ref{fig:whitening}).

\paragraph{Concept TF--IDF.} OpenAlex concepts at level $\geq 2$,
tokenized with pattern \texttt{[a-z\_]+}, \texttt{min\_df=10},
\texttt{max\_df=0.5}, then reduced to $128$ dimensions with
\texttt{TruncatedSVD(random\_state=42)}.

\paragraph{Venue LDA.} Linear Discriminant Analysis with venue slug as
the class label, yielding up to $(n_{\text{venues}}-1) = 28$ components.
Venues contributing fewer than $100$ papers at fit time are folded into
an ``other'' class, which is why the realized dimension is $28$ rather
than $33$.

\paragraph{Hybrid concatenation.} The three $L_2$-normalized blocks are
concatenated with square-root-weighted scaling: whitened SPECTER2 at
$\sqrt{0.55}$ ($768$ dim), concept TF--IDF at $\sqrt{0.30}$ ($128$ dim),
and venue LDA at $\sqrt{0.15}$ ($28$ dim). The $\sqrt{\cdot}$ scaling
makes the hybrid cosine an exact convex combination of the three block
cosines (Eq.~\ref{eq:cosine-decomposition}). The concatenated vector is
$L_2$-renormalized, so cosine and dot product coincide in downstream
kNN code.

\subsection{Graphs}
\label{subsec:app-c-graphs}

\paragraph{Coauthor graph.} Nodes are author keys after alias resolution
with at least two corpus papers
(\texttt{min\_papers\_per\_author = 2}). An edge $(a,b)$ carries weight
equal to the number of papers $a$ and $b$ jointly authored. The
displayed graph retains edges with weight $\geq 2$
(\texttt{BASE\_THRESHOLD = 2}, matching
\citet{sun2017coauthorship}), while the stricter giant-component
subgraph passed to Leiden retains edges with weight $\geq 5$
(\texttt{GIANT\_THRESHOLD = 5}). Connected components smaller than
$10$ nodes are merged into a miscellaneous-island bucket.

\paragraph{Semantic kNN graph (paper level).} Mutual
$k$-nearest-neighbor graph at $k{=}20$ in the hybrid cosine metric,
used as substrate for the paper-level semantic Leiden partition.

\paragraph{Author centroids.} Per-author citation-weighted mean of
hybrid paper vectors, restricted to authors with $\geq 2$ corpus papers
and non-trivial norm. Centroids are $L_2$-renormalized post-weighting.

\paragraph{Multiplex edges.} The multiplex graph is built on the
$\num{41551}$ authors of the coauthor graph. The coauthor layer
contributes edges with weight $\alpha \log(1 + w_{ab})$, where $w_{ab}$
is the number of papers $a$ and $b$ coauthored and $\alpha = 0.5$. The
semantic layer contributes an edge whenever $a$ and $b$ are
mutual top-$5$ author-centroid neighbours and the cosine similarity
exceeds $\tau_s = 0.60$; the semantic weight is the soft rescaling
$(1-\alpha)(\cos(a,b)-\tau_s)/(1-\tau_s)$, calibrated so that the
strongest semantic and coauthor ties have comparable magnitude.

\paragraph{Leiden.} \texttt{leidenalg} with
\texttt{RBConfigurationVertexPartition}. Resolution $1.0$ for both the
coauthor mainland and the paper-level semantic partition; resolution
$0.5$ for the combined multiplex. The Leiden RNG seed is $42$, and the
iteration budget is left at the library default (run to convergence).

\paragraph{ForceAtlas2.} \texttt{fa2\_modified} with
$500$ iterations and \texttt{scaling\_ratio}${=}2.0$. The layout is
precomputed in Python so that the browser never re-runs physics for
more than $5{,}000$ nodes.

\paragraph{UMAP (topic space).} \texttt{n\_neighbors}${=}15$,
\texttt{min\_dist}${=}0.1$, seed~$42$, cosine metric on
author centroids.

\subsection{Phantom-collaborator evaluation}
\label{subsec:app-c-phantom}

\paragraph{Splits.} Train cutoff $Y=2019$; hold-out window
$[2020, 2025]$. Both the whitening and the coauthor graph are refitted
on the pre-$2020$ subset so that no hold-out information leaks into the
predictor. The train subset contains $\num{73586}$ embedded papers and
yields $\num{27713}$ eligible authors, of whom $\num{5225}$ acquire at
least one realized coauthor in the hold-out window.

\paragraph{Predictor.} For each anchor $a$, the top-$K$ authors by
cosine similarity to the anchor's train-period centroid whose geodesic
distance in the train-period coauthor graph satisfies $d_{G^{\leq 2019}}(a,b) \geq 3$
(Eq.~\ref{eq:phantom-def}). The distance gate excludes direct coauthors
($d = 1$) and friends-of-friends ($d = 2$).

\paragraph{Evaluation grid.} $K \in \{5, 10, 20\}$. Headline lift
numbers are reported at $K=20$. The similarity-rate gradient
(Fig.~\ref{fig:phantom-calibration}) uses ten equal-frequency quantile
buckets of pairwise cosine similarity
over the top-$20$ phantom pairs.

\paragraph{Baselines.} Each anchor receives $K$ candidates drawn under
three null models, with direct- and near-coauthors ($d \leq 2$) excluded
at sampling time.
\begin{itemize}[leftmargin=*]
  \item \textbf{Random:} uniform over eligible authors.
  \item \textbf{Popularity-weighted:} probability proportional to the
        candidate's train-period paper count (a pragmatic
        configuration-model surrogate).
  \item \textbf{Same-venue:} uniform over candidates sharing at least
        one train-period publication venue with the anchor.
\end{itemize}

\subsection{Trajectories (Sec.~\ref{sec:trajectories})}
\label{subsec:app-c-traj}

Five-year bins, minimum two non-empty bins per author for
classification eligibility, restricted to authors with $\geq 3$ bins
($\geq\!10$-year publication window, two trajectory segments) for
the headline four-class partition. Centroid trajectories are smoothed
only through bin aggregation; no temporal regulariser or kernel
smoother is applied. Class cutoffs: stayer cutoff
$\tau_{\mathrm{stay}} = 15$, drifter cutoff $\tau_{\eta} = 0.60$,
switcher cutoff $\tau_{\mathrm{net}} = 15$ (all in UMAP units of the
Sec.~\ref{subsec:semantic-centroids} embedding).

\subsection{Random seeds}
\label{subsec:app-c-seeds}

The pipeline master seed is $1337$ (propagated to \texttt{numpy.random}
and the stdlib \texttt{random} module at entry of each stage).
Component-level overrides are \texttt{TruncatedSVD(random\_state=42)},
Leiden seed $42$, and UMAP seed $42$. ForceAtlas2 is deterministic
given the NetworkX edge-insertion order, which the pipeline fixes by
sorting edges lexicographically before layout. Re-running from the raw
OpenAlex snapshot reproduces every downstream artifact under
bit-identical input.

\clearpage
\section{Author disambiguation audit}
\label{app:disambig}

Author disambiguation has two automatic stages and a manually
curated override. The \textbf{resolution chain}
(Sec.~\ref{subsec:data-authors}) assigns each author-paper row to (i)
its OpenAlex author identifier when present, (ii) its ORCID when
present and the OpenAlex identifier is missing or spurious, and
(iii) a normalized surname-plus-first-initial key as final fallback.
The \textbf{automatic ORCID merger} then collapses OpenAlex
identifiers whose ORCIDs match under different canonical names; on
the current corpus this folds $\num{4543}$ stub identifiers into
$\num{3567}$ canonical identities. A \textbf{manually curated alias
list} handles the residual cases, typically researchers who hold two
distinct valid ORCIDs (duplicate registration across affiliation
portals) or who carry no ORCID on older papers. The list currently
contains $18$ canonical entries that consolidate $39$ OpenAlex
identifiers, drawn from the split-identity audit
(Sec.~\ref{subsec:disambig-split}). The dual-model LLM audit
(Sec.~\ref{subsec:disambig-llm}) additionally flagged $7$
high-confidence consensus pairs, of which manual review confirmed
$6$ as legitimate same-person merges; these are not yet folded into
the list pending the next pipeline rerun. Entries are admitted only
after passing one of the two audits below.

\subsection{LLM-based sanity check on borderline pairs}
\label{subsec:disambig-llm}

To assess whether the automatic stages leave high-confidence merges
on the table, we constructed a borderline set consisting of pairs
that (i) share a surname and a first initial, (ii) have at least $2$
shared coauthors in the pre-$2020$ coauthor graph, and (iii) are not
already merged by either automatic stage. The full population is
$\mathbf{372}$ pairs. We submitted every pair to two language models
independently (Anthropic Claude Haiku and Claude Sonnet), supplying
the two canonical names, ORCID status, top venues, three
representative paper titles, and shared coauthor surnames; each model
returned a three-way decision (same, different, unsure) with a
confidence score.

\Cref{tab:disambig-llm} reports the cross-model agreement. The two
models agree on $157$ of $372$ pairs ($42\%$), including $24$
``different'' verdicts and $96$ ``unsure'' verdicts on the unambiguous
ends of the population. They jointly flag $37$ pairs as the same
person at any confidence and $7$ at mutual confidence $\geq 0.8$.
Manual review of these seven confirms $\mathbf{6}$ as legitimate
name variants (\textit{Amditis, Angelos} vs \textit{Amditis,
Angelos~J.}, $18$ shared coauthors; \textit{Goo, Swee Keow} vs
\textit{Goo, Swee}; \textit{Johnson, Kent~A.} vs \textit{Johnson,
Kent}; \textit{Labeau, F.} vs \textit{Labeau, Fabrice}; \textit{Su,
Kui} vs \textit{Su, Kuifeng}; \textit{Crandall, Jeff~R.} vs
\textit{Crandall, Jeff}). The seventh, \textit{Kim, Jeongheon} vs
\textit{Kim, Jaeik}, is a false positive: the two given names are
distinct Korean given names whose carriers share a research-lab
footprint, and both models were fooled by the shared-coauthor signal
without recognizing the given-name divergence. This reinforces that
LLM consensus is a useful triage filter but not a substitute for
human review on common-surname cases. The remaining single-model
``same'' flags ($117$ from Sonnet not endorsed by Haiku, $5$ in the
reverse direction) sit in the borderline zone and are held back
pending stronger evidence.

\begin{table}[pos = h, width=0.85\textwidth]
  \centering
  \small
  \caption{Cross-model agreement on $372$ borderline author-alias
    candidates (Claude Haiku rows $\times$ Claude Sonnet columns).}
  \label{tab:disambig-llm}
  \begin{tabular}{lrrrr}
    \toprule
                       & \multicolumn{3}{c}{\textbf{Sonnet}} &  \\
    \cmidrule(lr){2-4}
    \textbf{Haiku}     & \textbf{same} & \textbf{different} & \textbf{unsure} & \textbf{Total} \\
    \midrule
    same               & $37$   & $0$    & $5$    & $42$  \\
    different          & $11$   & $24$   & $10$   & $45$  \\
    unsure             & $106$  & $83$   & $96$   & $285$ \\
    \midrule
    \textbf{Total}     & $154$  & $107$  & $111$  & $372$ \\
    \bottomrule
  \end{tabular}
\end{table}

\subsection{Split-identity audit}
\label{subsec:disambig-split}

A complementary audit targets the opposite failure mode: identity
\emph{splits}, where OpenAlex has assigned two distinct identifiers
to the same researcher. For every same-name group spanning two or
more OpenAlex identifiers, in which at least one identifier holds
$\geq 5$ papers, we compute pairwise shared-coauthor counts and the
set of coauthor edges that would be \emph{restored} if the pair were
merged --- third-party coauthors whose combined pair count with both
identifiers clears the display edge threshold but for whom no edge
currently attaches to either identifier. We then group candidate
pairs by confidence using four signals: ORCID asymmetry,
institutional overlap, shared-coauthor count, and same-Leiden-community
membership. \Cref{tab:disambig-split} reports the resulting
breakdown over the $\num{163008}$ OpenAlex-identified authors.

\begin{table}[pos = h, width=0.95\textwidth]
  \centering
  \small
  \caption{Split-identity audit over the pre-alias coauthor graph.}
  \label{tab:disambig-split}
  \begin{tabular}{lrr}
    \toprule
    \textbf{Confidence group} & \textbf{Candidate pairs} & \textbf{Dropped edges restored} \\
    \midrule
    High: one identifier without ORCID, $\geq 3$ shared coauthors, shared institution & $13$ & $8$ \\
    Ambiguous-high: $\geq 20$ shared coauthors, same institution, different ORCIDs & $2$ & $10$ \\
    Moderate: $\geq 3$ shared coauthors, partial institutional overlap & $58$ & $55$ \\
    Weak: exactly $2$ shared coauthors, both ORCIDs present & $91$ & $50$ \\
    Confirmed false positives & $2$ & -- \\
    \midrule
    \textbf{Total non-FP}  & \textbf{164} & \textbf{123} \\
    \bottomrule
  \end{tabular}
\end{table}

Against the $\num{83035}$ edges in the current coauthor graph, a
worst-case merge of every audited pair would add at most $123$ edges
relative to the pre-alias baseline, an upper bound of $0.15\%$ on
split-identity noise. We applied $14$ new merges manually --- the
$13$ high-confidence pairs plus the one ambiguous-high pair
(\textit{Zhou, Xuesong}) that survived a manual cross-check against
OpenAlex --- on top of the $4$ pre-existing entries in the alias
list. The other ambiguous-high candidate (\textit{Chen, Long}) is
deliberately \emph{not} merged: OpenAlex's primary record for this
name already conflates multiple distinct researchers globally (over
$490$ works spanning more than $120$ institutions), so merging a
second identifier into it would worsen rather than repair identity
resolution, and we defer that case to a future split-then-merge pass.
The moderate- and weak-confidence groups are held back pending either
ORCID backfill or a stricter rule, because spot-checks in those
groups include genuine homonyms at shared institutions.

\subsection{Patterns and residual failure modes}
\label{subsec:disambig-patterns}

The audit surfaces four recurrent mechanisms behind the remaining
splits. First, the dominant pattern is \textbf{mid-career ORCID
registration}: an author publishes without ORCID for several years,
then registers one; OpenAlex creates a new identifier against the
ORCID and fails to back-link earlier papers. Among the high-confidence
and ambiguous-high pairs, $13$ of $15$ have exactly one identifier
without ORCID, with the ORCID-bearing identifier's first year
clustering in $2016$--$2021$ (the post-ORCID adoption wave in Chinese
universities and transport journals). Second, a handful of researchers
hold \textbf{duplicate ORCID registrations} from successive affiliation
portals; the already-merged \textit{Wu, Yuankai} and \textit{Sun,
Lijun} (McGill) entries are confirmed instances. Third, the
unclaimed-ORCID prefix introduced in $2022$ is a useful diagnostic:
ten high- and moderate-confidence pairs carry one such prefix
alongside a normal ORCID for the same name and institution, a
reliable fingerprint of a portal-generated duplicate. Fourth,
candidate splits cluster in the three largest venues --- T-ITS
($79$ occurrences), TR-C ($65$), and TRR ($57$) --- a size effect
compounded by the fact that multi-venue authors who publish across
the IEEE and Elsevier ecosystems (which index separately) are at
higher risk of fragmented identifiers.

The dominant \textbf{residual failure mode} is homonymy among common
East Asian surnames without ORCID anchoring. The two confirmed false
positives in the high-confidence audit both involve \textit{Sun, Lijun}
at Tongji University, where three senior transport faculty co-publish
across pavement, traffic-flow, and control subfields; their broad
collaborative footprint produces $3$--$4$ shared coauthors for
genuinely distinct researchers who share a common Chinese name. The
requirement that exactly one of the two identifiers lacks an ORCID is
therefore a necessary precondition for any automated expansion beyond
the high-confidence cohort. This bias affects the long tail of the
productivity distribution rather than the top contributors analyzed in
Sec.~\ref{sec:descriptive}, where spot-checks of the top-$30$
centrality ranks found no misattributed records.


\clearpage
\section{Extended descriptive bibliometrics}
\label{app:descriptive-extended}

This appendix supplements Sec.~\ref{sec:descriptive} with per-venue
bibliometric detail and the long-tail productivity distribution. These
tables and figures are reference material: they document who publishes
where and how often, but do not feed into the coauthor-network,
topic-space, or phantom-collaborator results that form the paper's
central contributions.

\subsection{Per-venue bibliometric summary}
\label{app:desc-venue-stats}

\Cref{tab:venue-stats} extends \citet{sun2017coauthorship}'s Table~2 to
our 34-venue, 60-year corpus, computed across all paper types
(counts run slightly higher than \Cref{tab:app-venues}, which
restricts to OpenAlex \texttt{journal-article} and
\texttt{conference-paper} types). Four observations. TRR is the
volume leader with $\num{38628}$ papers. IEEE T-ITS has the largest average
team size in the post-$2015$ era. TR-F has the smallest single-author
fraction despite its behavioral-science focus. Transportation Science
(INFORMS) has the highest mean citations per paper among the
operations-research venues.

\begin{table}[width=0.99\textwidth, pos = h]
  \centering
  \small
  \caption{Per-venue bibliometric summary.}
  \label{tab:venue-stats}
    \begin{tabular}{lrrrrrrrr}
  \toprule
  \textbf{Venue} & \textbf{Papers} & \textbf{Authors} & \textbf{Single} & \textbf{Avg} & \textbf{Max} & \textbf{Collabs} & \textbf{P/A} & \textbf{Cites} \\
                  & \textbf{} & \textbf{} & \textbf{\%} & \textbf{auth.} & \textbf{auth.} & \textbf{} & \textbf{} & \textbf{/paper} \\
  \midrule
  TR-A      &  5,131 &  9,384 & 11.9 & 2.84 & 73 &  28,584 & 0.55 & 49.3 \\
  TR-B      &  3,827 &  4,731 & 16.5 & 2.60 & 12 &  11,462 & 0.81 & 71.8 \\
  TR-C      &  4,885 &  9,865 &  3.8 & 3.59 & 16 &  28,881 & 0.50 & 62.9 \\
  TR-D      &  4,668 & 12,447 &  6.2 & 3.67 & 26 &  30,899 & 0.38 & 48.2 \\
  TR-E      &  4,021 &  8,437 &  5.8 & 3.36 & 15 &  19,672 & 0.48 & 50.9 \\
  TR-F      &  3,556 &  8,143 &  5.3 & 3.78 & 35 &  25,903 & 0.44 & 38.4 \\
  T-ITS     & 11,368 & 31,250 &  3.1 & 4.36 & 97 & 118,858 & 0.36 & 43.9 \\
  T-IV      &  2,165 &  6,534 &  1.4 & 4.38 & 20 &  21,937 & 0.33 & 26.8 \\
  ITS Mag   &  1,496 &  3,153 & 25.6 & 2.81 & 65 &  15,214 & 0.47 & 17.9 \\
  CommTR    &    204 &    779 &  2.5 & 4.56 & 15 &   2,192 & 0.26 & 20.6 \\
  TransSci  &  2,570 &  3,407 & 18.8 & 2.40 & 16 &   6,734 & 0.75 & 64.3 \\
  TRR       & 38,628 & 47,217 & 16.0 & 2.83 & 24 & 140,372 & 0.82 & 15.6 \\
  AMAR      &    280 &    494 &  2.1 & 3.64 & 13 &   1,684 & 0.57 & 69.6 \\
  TranspRev &  1,767 &  2,587 & 34.7 & 2.09 & 30 &   4,792 & 0.68 & 48.9 \\
  JTG       &  4,225 &  7,261 & 26.7 & 2.67 & 29 &  15,589 & 0.58 & 45.7 \\
  AAP       &  9,324 & 16,774 & 13.0 & 3.30 & 31 &  56,655 & 0.56 & 50.5 \\
  JUM       &    207 &    647 &  5.8 & 3.50 &  9 &   1,157 & 0.32 &  9.8 \\
  TBS       &  1,178 &  3,294 &  5.0 & 3.57 & 22 &   7,121 & 0.36 & 23.8 \\
  TranspPol &  3,722 &  7,950 & 14.1 & 3.00 & 17 &  15,842 & 0.47 & 43.3 \\
  RTBM      &  1,312 &  3,460 & 10.4 & 3.14 & 10 &   5,808 & 0.38 & 20.3 \\
  ETRR      &    774 &  2,045 &  9.2 & 3.32 & 14 &   4,249 & 0.38 & 27.4 \\
  IJST      &  1,014 &  2,681 &  8.9 & 3.24 & 15 &   4,920 & 0.38 & 30.2 \\
  TRIP      &  1,834 &  5,544 &  8.6 & 3.60 & 18 &  12,340 & 0.33 & 16.0 \\
  JPT       &    682 &  1,536 & 19.5 & 2.76 & 54 &   3,639 & 0.44 & 29.1 \\
  CACIE     &  2,871 &  6,301 &  8.5 & 3.12 & 30 &  17,591 & 0.46 & 33.7 \\
  JICV      &    147 &    616 &  1.4 & 4.60 & 13 &   1,470 & 0.24 & 15.0 \\
  JTTE      &    722 &  2,347 &  3.7 & 3.88 & 43 &   6,057 & 0.31 & 26.8 \\
  RailEng   &    229 &    859 &  3.1 & 4.76 & 13 &   2,513 & 0.27 & 15.9 \\
  OJ-ITS    &    425 &  1,502 &  2.8 & 3.97 & 25 &   3,333 & 0.28 & 10.7 \\
  IJTST     &    696 &  1,891 &  5.0 & 3.55 & 11 &   4,014 & 0.37 & 17.2 \\
  eTransp   &    507 &  2,479 &  1.6 & 6.26 & 35 &  11,099 & 0.20 & 45.9 \\
  GEIT      &    235 &  1,103 &  2.6 & 5.23 & 15 &   3,282 & 0.21 & 27.5 \\
  T-TE      &  4,038 & 11,973 &  0.5 & 4.99 & 62 &  48,718 & 0.34 & 23.8 \\
  VT Mag    &  1,615 &  3,041 & 30.5 & 3.05 & 60 &  32,241 & 0.53 & 25.6 \\
  \bottomrule
  \end{tabular}
\end{table}

\subsection{Top contributors and most-cited papers}
\label{app:desc-top}

\Cref{tab:top-contributors} lists the top-five contributors per venue
by raw paper count and by total citations earned in that venue.
\Cref{tab:top-papers} shows the three most-cited papers per venue,
a who's-who of landmark transportation studies.

\begin{landscape}
    
\begin{table}
  \centering
  \footnotesize
  \caption{Top-five contributors per venue.}
  \label{tab:top-contributors}
  \resizebox{1.35\textwidth}{!}{
    \begin{tabular}{lll}
  \toprule
  \textbf{Venue} & \textbf{Top-5 contributors (papers)} & \textbf{Top-5 contributors (citations)} \\
  \midrule
  TR-A & David A. Hensher (74); Khandker Nurul Habib (39); Anming Zhang (33); José Holguín-Veras (30); Kay W. Axhausen (29) & David A. Hensher (5,924); Kara M. Kockelman (4,368); Daniel J. Fagnant (3,038); Patricia L. Mokhtarian (2,942); Jillian Anable (2,585) \\
  TR-B & Hai Yang (108); Carlos F. Daganzo (77); Chandra R. Bhat (70); Qiang Meng (62); S.C. Wong (53) & Carlos F. Daganzo (15,730); Hai Yang (12,408); Chandra R. Bhat (8,997); Nikolas Geroliminis (5,594); Gilbert Laporte (5,207) \\
  TR-C & Hai Yang (66); Markos Papageorgiou (44); Yafeng Yin (41); Serge Hoogendoorn (41); Ziyou Gao (39) & Yinhai Wang (5,133); Markos Papageorgiou (4,638); Hani S. Mahmassani (4,084); Hai Yang (4,071); Serge Hoogendoorn (3,733) \\
  TR-D & Xinyu Cao (24); Jonn Axsen (23); Marianne Hatzopoulou (21); Guohua Song (19); H. Oliver Gao (17) & Robert Cervero (5,197); Kara M. Kockelman (4,743); Susan Handy (2,411); Xinyu Cao (2,376); Patricia L. Mokhtarian (2,363) \\
  TR-E & Jiuh-Biing Sheu (50); Qiang Meng (45); Shuaian Wang (43); T.C.E. Cheng (40); Tsan-Ming Choi (36) & Tsan-Ming Choi (6,269); Dmitry Ivanov (4,712); Jiuh-Biing Sheu (4,671); Qiang Meng (3,375); Kannan Govindan (3,224) \\
  TR-F & Türker Özkan (44); Joost de Winter (40); Natasha Merat (39); Óscar Oviedo-Trespalacios (39); Timo Lajunen (38) & Joost de Winter (4,436); Riender Happee (4,079); Timo Lajunen (3,269); Josef F. Krems (2,915); Natasha Merat (2,794) \\
  T-ITS & Fei-Yue Wang (75); Azim Eskandarian (74); MengChu Zhou (73); Pétros Ioannou (62); Tao Tang (47) & Fei-Yue Wang (10,319); Mohan M. Trivedi (7,204); Tao Tang (4,934); Yisheng Lv (4,622); MengChu Zhou (4,282) \\
  T-IV & Fei-Yue Wang (111); Yutong Wang (30); Xiao Wang (24); Yonglin Tian (22); Hong Chen (22) & Fei-Yue Wang (4,050); Sze Zheng Yong (2,533); Michal \v{C}\'{a}p (2,464); Emilio Frazzoli (2,464); B. Paden (2,464) \\
  ITS Mag & Yisheng Lv (48); Ljubo Vlacic (45); Cristina Olaverri-Monreal (45); Martin Lauer (44); Miguel Ángel Sotelo (34) & Christoph Stiller (1,805); Cyrill Stachniss (1,293); Wolfram Burgard (1,293); Rainer Kümmerle (1,293); Giorgio Grisetti (1,293) \\
  CommTR & Xiaobo Qu (15); Shuaian Wang (9); Balázs Kulcsár (6); Sikai Chen (6); Zuduo Zheng (5) & Xiaobo Qu (588); Shuaian Wang (424); Yang Liu (333); Balázs Kulcsár (233); Zuduo Zheng (219) \\
  TransSci & Martin Savelsbergh (40); Warren B. Powell (33); Gilbert Laporte (33); Michel Gendreau (32); Carlos F. Daganzo (31) & Michel Gendreau (6,087); Gilbert Laporte (5,659); Martin Savelsbergh (4,565); Mark S. Daskin (3,537); Stefan Røpke (3,295) \\
  TRR & Hani S. Mahmassani (226); Imad L. Al-Qadi (145); Nagui M. Rouphail (135); Serge Hoogendoorn (129); Mohamed Abdel-Aty (121) & Hani S. Mahmassani (7,262); Chandra R. Bhat (5,203); Mohamed Abdel-Aty (4,937); Serge Hoogendoorn (4,771); Moshe Ben-Akiva (4,667) \\
  AMAR & Fred Mannering (39); Md. Mazharul Haque (34); Naveen Eluru (25); Panagiotis Ch. Anastasopoulos (22); Shamsunnahar Yasmin (20) & Fred Mannering (7,263); Chandra R. Bhat (3,100); Panagiotis Ch. Anastasopoulos (1,964); Ali Behnood (1,789); Venky Shankar (1,786) \\
  TranspRev & David A. Hensher (38); David Banister (36); Bert van Wee (28); Kenneth Button (17); Graham Currie (15) & Bert van Wee (4,639); Ralph Buehler (3,896); John Pucher (3,429); David Banister (2,433); Susan Handy (1,960) \\
  JTG & Frank Witlox (41); Richard Knowles (36); Antonio Páez (26); Darren M. Scott (22); Graham Currie (22) & Bert van Wee (3,792); Karst Geurs (3,207); Antonio Páez (3,095); Frank Witlox (2,876); David Banister (2,696) \\
  AAP & Mohamed Abdel-Aty (150); Rune Elvik (106); Helai Huang (61); Tarek Sayed (60); Dominique Lord (56) & Mohamed Abdel-Aty (13,060); Fred Mannering (9,386); Helai Huang (5,579); Rune Elvik (5,408); Mohammed Quddus (4,563) \\
  JUM & Benjamin Büttner (5); Filipe Moura (4); Karst Geurs (4); Domokos Esztergár-Kiss (4); Ulrike Jehle (3) & Benjamin Büttner (150); Karst Geurs (146); Luca Staricco (114); Ehab Diab (94); Clément Lemardelé (90) \\
  TBS & Long Cheng (17); Jonas De Vos (15); Dick Ettema (12); Linchuan Yang (12); Soora Rasouli (9) & Susan Handy (1,027); Jonas De Vos (994); Frank Witlox (724); Patricia L. Mokhtarian (675); Long Cheng (668) \\
  TranspPol & David A. Hensher (41); Kun Wang (28); Xiaowen Fu (25); Anming Zhang (21); Tom Rye (18) & David Banister (3,832); Robert Cervero (3,658); David A. Hensher (2,558); Karen Lucas (2,336); Tommy Gärling (1,877) \\
  RTBM & Thierry Vanelslander (13); John D. Nelson (11); Stephen Ison (10); Theo Notteboom (10); Jason Monios (8) & John D. Nelson (539); Thierry Vanelslander (499); Miriam Ricci (455); Theo Notteboom (446); Lætitia Dablanc (346) \\
  ETRR & Cathy Macharis (11); Dewan Md Zahurul Islam (9); Elisabete Arsénio (9); George Yannis (8); Constantinos Antoniou (8) & Lelitha Vanajakshi (784); S. Vasantha Kumar (784); Cathy Macharis (577); Biagio Ciuffo (401); Constantinos Antoniou (396) \\
  IJST & Keechoo Choi (20); Asad J. Khattak (12); Margarida C. Coelho (10); Robert B. Noland (8); Satoshi Fujii (8) & Susan Shaheen (762); Susan Handy (687); Keechoo Choi (595); Adam Cohen (550); Ahmed El-Geneidy (536) \\
  TRIP & Karl Kim (11); John D. Nelson (9); Marjan Hagenzieker (8); Prawira Fajarindra Belgiawan (8); Sharareh Kermanshachi (8) & Jonas De Vos (1,123); Marije Hamersma (765); Muhammad Abdullah (745); Charitha Dias (745); Roel Faber (741) \\
  JPT & Jeffrey Brown (10); Graham Currie (9); Avishai Ceder (8); Albert Gan (7); Oded Cats (7) & Paul DeMaio (1,197); Gabriella Mazzulla (905); Laura Eboli (905); Oded Cats (833); Alejandro Tirachini (791) \\
  CACIE & Hojjat Adeli (54); S. Travis Waller (27); Dawei Wang (24); Paul Schonfeld (23); D Loucks (23) & Young-Jin Cha (5,140); Oral Büyüköztürk (4,793); Wooram Choi (4,555); Hojjat Adeli (3,837); Kelvin C. P. Wang (1,616) \\
  JICV & Kun Gao (4); Changxi Ma (4); Jiaming Wu (3); Lingxi Li (3); Rencheng Zheng (3) & Kun Gao (207); Said M. Easa (129); Jiaming Wu (117); Jie Zhu (115); Xiaobo Qu (103) \\
  JTTE & Zhanping You (11); Khaled Ksaibati (11); Mohd Rosli Mohd Hasan (10); Yongjian Liu (9); Bruno Briseghella (8) & Weixing Wang (692); Zhanping You (687); Peiliang Cong (536); Mohd Rosli Mohd Hasan (471); Yaqian Cheng (457) \\
  RailEng & Wanming Zhai (13); Maksym Spiryagin (11); Colin Cole (8); Qing Wu (7); Lei Xu (6) & Wanming Zhai (344); Maksym Spiryagin (231); Sebastian Stichel (205); Lei Xu (203); Zefeng Wen (200) \\
  OJ-ITS & Bart van Arem (8); Frank Gauterin (6); Michael Frey (6); Shaurya Agarwal (6); Francesco Corman (5) & Rodolfo Valiente (113); Shaurya Agarwal (108); Seonwook Kim (105); Yonghwan Jeong (105); Kyongsu Yi (105) \\
  IJTST & Zhongren Wang (22); Hesham Rakha (14); Hui Li (13); Subasish Das (12); Chiu Liu (12) & Hesham Rakha (455); Wei Fan (447); Eleni I. Vlahogianni (444); Hui Li (373); John Golias (338) \\
  eTransp & Minggao Ouyang (28); Xuning Feng (17); Dirk Uwe Sauer (14); Languang Lu (13); Markus Lienkamp (13) & Minggao Ouyang (4,927); Xuning Feng (4,369); Yuejiu Zheng (2,894); Languang Lu (2,819); Xuebing Han (2,576) \\
  GEIT & Rui Xiong (6); Yonggang Liu (5); Chen Lv (4); Quanqing Yu (4); Junfu Li (4) & Rui Xiong (674); Fengchun Sun (511); Zhongbao Wei (395); Hongwen He (382); Zhenpo Wang (313) \\
  T-TE & Xiaodong Sun (42); Ali Emadi (41); Zhuoran Zhang (41); Wei Hua (40); Zhigang Liu (35) & Ali Emadi (2,927); Bulent Sarlioglu (2,109); Xiaodong Sun (1,890); Babak Fahimi (1,787); Berker Bilgin (1,566) \\
  VT Mag & Javier Gozálvez (73); Harvey Glickenstein (60); Bill Fleming (38); Elisabeth Uhlemann (34); Klaus David (33) & Zhiguo Ding (2,508); Zhengquan Zhang (2,482); George K. Karagiannidis (2,482); Pingzhi Fan (2,480); Zheng Ma (2,477) \\
  \bottomrule
  \end{tabular}
  }
\end{table}

\end{landscape}

\begin{landscape}
  \footnotesize
  \begin{longtable}{lp{14cm}rrp{4.5cm}}
  \caption{Three most-cited papers per venue.}
  \label{tab:top-papers} \\
  \toprule
  \textbf{Venue} & \textbf{Title} & \textbf{Year} & \textbf{Cites} & \textbf{First author} \\
  \midrule
  \endfirsthead
  \multicolumn{5}{l}{\emph{\tablename~\thetable{} -- continued from previous page}} \\
  \toprule
  \textbf{Venue} & \textbf{Title} & \textbf{Year} & \textbf{Cites} & \textbf{First author} \\
  \midrule
  \endhead
  \midrule
  \multicolumn{5}{r}{\emph{continued on next page}} \\
  \endfoot
  \bottomrule
  \endlastfoot
  TR-A & Preparing a nation for autonomous vehicles: opportunities, barriers and policy recommendations & 2015 & 3,038 & Daniel J. Fagnant \\
   & The statistical analysis of crash-frequency data: A review and assessment of methodological alternatives & 2010 & 1,670 & Dominique Lord \\
   & Car use: lust and must. Instrumental, symbolic and affective motives for car use & 2004 & 1,267 & Linda Steg \\
  \addlinespace
  TR-B & The cell transmission model: A dynamic representation of highway traffic consistent with the hydrodynamic theory & 1994 & 2,863 & Carlos F. Daganzo \\
   & A behavioural car-following model for computer simulation & 1981 & 2,305 & P G Gipps \\
   & The cell transmission model, part II: Network traffic & 1995 & 1,941 & Carlos F. Daganzo \\
  \addlinespace
  TR-C & Long short-term memory neural network for traffic speed prediction using remote microwave sensor data & 2015 & 2,109 & Xiaolei Ma \\
   & The flying sidekick traveling salesman problem: Optimization of drone-assisted parcel delivery & 2015 & 1,518 & Chase Murray \\
   & Influence of connected and autonomous vehicles on traffic flow stability and throughput & 2016 & 1,314 & Alireza Talebpour \\
  \addlinespace
  TR-D & Travel demand and the 3Ds: Density, diversity, and design & 1997 & 4,342 & Robert Cervero \\
   & Advances in consumer electric vehicle adoption research: A review and research agenda & 2014 & 1,200 & Zeinab Rezvani \\
   & Correlation or causality between the built environment and travel behavior? Evidence from Northern California & 2005 & 1,118 & Susan Handy \\
  \addlinespace
  TR-E & Predicting the impacts of epidemic outbreaks on global supply chains: A simulation-based analysis on the coronavirus outbreak (COVID-19/SARS-CoV-2) case & 2020 & 2,139 & Dmitry Ivanov \\
   & Blockchain technology in supply chain operations: Applications, challenges and research opportunities & 2020 & 1,452 & Pankaj Dutta \\
   & The influence of greening the suppliers and green innovation on environmental performance and competitive advantage in Taiwan & 2011 & 1,283 & T.-Y. Chiou \\
  \addlinespace
  TR-F & Public opinion on automated driving: Results of an international questionnaire among 5000 respondents & 2015 & 1,300 & Miltos Kyriakidis \\
   & Car-following: a historical review & 1999 & 1,262 & Mark Brackstone \\
   & Effects of visual and cognitive load in real and simulated motorway driving & 2005 & 813 & Johan Engström \\
  \addlinespace
  T-ITS & T-GCN: A Temporal Graph Convolutional Network for Traffic Prediction & 2019 & 3,062 & Ling Zhao \\
   & Traffic Flow Prediction With Big Data: A Deep Learning Approach & 2014 & 2,948 & Yisheng Lv \\
   & Detecting Stress During Real-World Driving Tasks Using Physiological Sensors & 2005 & 2,088 & Jennifer Healey \\
  \addlinespace
  T-IV & A Survey of Motion Planning and Control Techniques for Self-Driving Urban Vehicles & 2016 & 2,464 & B. Paden \\
   & Simultaneous Localization and Mapping: A Survey of Current Trends in Autonomous Driving & 2017 & 860 & Guillaume Bresson \\
   & A Survey on Trajectory-Prediction Methods for Autonomous Driving & 2022 & 586 & Yanjun Huang \\
  \addlinespace
  ITS Mag & A Tutorial on Graph-Based SLAM & 2010 & 1,293 & Giorgio Grisetti \\
   & Three Decades of Driver Assistance Systems: Review and Future Perspectives & 2014 & 849 & Klaus Bengler \\
   & Making Bertha Drive---An Autonomous Journey on a Historic Route & 2014 & 843 & Julius Ziegler \\
  \addlinespace
  CommTR & Trip energy consumption estimation for electric buses & 2022 & 137 & Jinhua Ji \\
   & Connected autonomous vehicles for improving mixed traffic efficiency in unsignalized intersections with deep reinforcement learning & 2021 & 125 & Bile Peng \\
   & How machine learning informs ride-hailing services: A survey & 2022 & 115 & Yang Liu \\
  \addlinespace
  TransSci & An Adaptive Large Neighborhood Search Heuristic for the Pickup and Delivery Problem with Time Windows & 2006 & 2,311 & Stefan Røpke \\
   & Self-Organized Pedestrian Crowd Dynamics: Experiments, Simulations, and Design Solutions & 2005 & 1,463 & Dirk Helbing \\
   & Network Design and Transportation Planning: Models and Algorithms & 1984 & 1,279 & Thomas L. Magnanti \\
  \addlinespace
  TRR & MODELING THE CHOICE OF RESIDENTIAL LOCATION & 1978 & 2,372 & Daniel McFadden \\
   & Travel and the Built Environment: A Synthesis & 2001 & 1,837 & Reid Ewing \\
   & General Lane-Changing Model MOBIL for Car-Following Models & 2007 & 1,278 & Arne Kesting \\
  \addlinespace
  AMAR & Unobserved heterogeneity and the statistical analysis of highway accident data & 2016 & 1,226 & Fred Mannering \\
   & Analytic methods in accident research: Methodological frontier and future directions & 2013 & 1,141 & Fred Mannering \\
   & Temporal instability and the analysis of highway accident data & 2017 & 449 & Fred Mannering \\
  \addlinespace
  TranspRev & Making Cycling Irresistible: Lessons from The Netherlands, Denmark and Germany & 2008 & 1,557 & John Pucher \\
   & Commuting by Bicycle: An Overview of the Literature & 2009 & 1,180 & Eva Heinen \\
   & Examining the Impacts of Residential Self-Selection on Travel Behaviour: A Focus on Empirical Findings & 2009 & 1,050 & Xinyu Cao \\
  \addlinespace
  JTG & Accessibility evaluation of land-use and transport strategies: review and research directions & 2003 & 3,019 & Karst Geurs \\
   & From roadkill to road ecology: A review of the ecological effects of roads & 2007 & 1,142 & Alisa W. Coffin \\
   & Transport and climate change: a review & 2007 & 1,109 & Lee Chapman \\
  \addlinespace
  AAP & Driving speed and the risk of road crashes: A review & 2005 & 1,239 & L T Aarts \\
   & Thirty years of safety climate research: Reflections and future directions & 2010 & 1,076 & Dov Zohar \\
   & Toward safer highways, application of XGBoost and SHAP for real-time accident detection and feature analysis & 2019 & 970 & Amir Bahador Parsa \\
  \addlinespace
  JUM & 15-, 10- or 5-minute city? A focus on accessibility to services in Turin, Italy & 2022 & 114 & Luca Staricco \\
   & A composite X-minute city cycling accessibility metric and its role in assessing spatial and socioeconomic inequalities -- A case study in Utrecht, the Netherlan & 2022 & 90 & Elizabeth Knap \\
   & Understanding the determinants of x-minute city policies: A review of the North American and Australian cities' planning documents & 2022 & 78 & Michael T. Lu \\
  \addlinespace
  TBS & What influences travelers to use Uber? Exploring the factors affecting the adoption of on-demand ride services in California & 2018 & 510 & Farzad Alemi \\
   & Applying a random forest method approach to model travel mode choice behavior & 2018 & 461 & Long Cheng \\
   & Prediction and behavioral analysis of travel mode choice: A comparison of machine learning and logit models & 2020 & 379 & Xilei Zhao \\
  \addlinespace
  TranspPol & The sustainable mobility paradigm & 2007 & 2,516 & David Banister \\
   & Transport and social exclusion: Where are we now? & 2012 & 1,673 & Karen Lucas \\
   & Towards a theory of decoupling: degrees of decoupling in the EU and the case of road traffic in Finland between 1970 and 2001 & 2005 & 1,587 & Petri Tapio \\
  \addlinespace
  RTBM & Bike sharing: A review of evidence on impacts and processes of implementation and operation & 2015 & 455 & Miriam Ricci \\
   & Final deliveries for online shopping: The deployment of pickup point networks in urban and suburban areas & 2014 & 303 & Eléonora Morganti \\
   & Total cost of ownership and its potential implications for battery electric vehicle diffusion & 2016 & 285 & Jens Hagman \\
  \addlinespace
  ETRR & Short-term traffic flow prediction using seasonal ARIMA model with limited input data & 2015 & 784 & S. Vasantha Kumar \\
   & Vision-based vehicle detection and counting system using deep learning in highway scenes & 2019 & 383 & Huansheng Song \\
   & Crowd logistics: an opportunity for more sustainable urban freight transport? & 2017 & 304 & Heleen Buldeo \\
  \addlinespace
  IJST & Carsharing and Personal Vehicle Services: Worldwide Market Developments and Emerging Trends & 2012 & 550 & Susan Shaheen \\
   & Understanding the usage of dockless bike sharing in Singapore & 2018 & 419 & Yu Shen \\
   & Factors of electric vehicle adoption: A comparison of conventional and electric car users based on an extended theory of planned behavior & 2018 & 306 & Sonja Haustein \\
  \addlinespace
  TRIP & The effect of COVID-19 and subsequent social distancing on travel behavior & 2020 & 971 & Jonas De Vos \\
   & How COVID-19 and the Dutch `intelligent lockdown' change activities, work and travel behaviour: Evidence from longitudinal data in the Netherlands & 2020 & 741 & Mathijs de Haas \\
   & Exploring the impacts of COVID-19 on travel behavior and mode preferences & 2020 & 681 & Muhammad Abdullah \\
  \addlinespace
  JPT & Bike-sharing: History, Impacts, Models of Provision, and Future & 2009 & 1,056 & Paul DeMaio \\
   & COVID-19 and Public Transportation: Current Assessment, Prospects, and Research Needs & 2020 & 791 & Alejandro Tirachini \\
   & Service Quality Attributes Affecting Customer Satisfaction for Bus Transit & 2007 & 572 & Laura Eboli \\
  \addlinespace
  CACIE & Deep Learning-Based Crack Damage Detection Using Convolutional Neural Networks & 2017 & 3,054 & Young-Jin Cha \\
   & Autonomous Structural Visual Inspection Using Region-Based Deep Learning for Detecting Multiple Damage Types & 2017 & 1,501 & Young-Jin Cha \\
   & Automated Pixel-Level Pavement Crack Detection on 3D Asphalt Surfaces Using a Deep-Learning Network & 2017 & 928 & Allen Zhang \\
  \addlinespace
  JICV & Merging control strategies of connected and autonomous vehicles at freeway on-ramps: a comprehensive review & 2022 & 115 & Jie Zhu \\
   & Dynamic prediction of traffic incident duration on urban expressways: a deep learning approach based on LSTM and MLP & 2021 & 85 & Weiwei Zhu \\
   & Intersection control with connected and automated vehicles: a review & 2022 & 80 & Jiaming Wu \\
  \addlinespace
  JTTE & Advances in geopolymer materials: A comprehensive review & 2021 & 457 & Peiliang Cong \\
   & A review of road extraction from remote sensing images & 2016 & 295 & Weixing Wang \\
   & Recent advances in connected and automated vehicles & 2019 & 246 & David L. Elliott \\
  \addlinespace
  RailEng & Polygonisation of railway wheels: a critical review & 2020 & 163 & Gongquan Tao \\
   & Review of research on high-speed railway subgrade settlement in soft soil area & 2020 & 136 & Shunhua Zhou \\
   & Active suspension in railway vehicles: a literature survey & 2020 & 133 & Bin Fu \\
  \addlinespace
  OJ-ITS & Surround Vehicle Motion Prediction Using LSTM-RNN for Motion Planning of Autonomous Vehicles at Multi-Lane Turn Intersections & 2020 & 105 & Yonghwan Jeong \\
   & Vehicle Classification in Intelligent Transport Systems: An Overview, Methods and Software Perspective & 2021 & 98 & Ashkan Gholamhosseinian \\
   & High-Definition Maps: Comprehensive Survey, Challenges, and Future Perspectives & 2023 & 93 & Gamal Elghazaly \\
  \addlinespace
  IJTST & Assessing the impacts of deploying a shared self-driving urban mobility system: An agent-based model applied to the city of Lisbon, Portugal & 2017 & 309 & Luis Martínez \\
   & Advancements, prospects, and impacts of automated driving systems & 2017 & 266 & Ching-Yao Chan \\
   & Unmanned Aerial Aircraft Systems for transportation engineering: Current practice and future challenges & 2016 & 219 & Emmanouil Barmpounakis \\
  \addlinespace
  eTransp & A review on the key issues of the lithium ion battery degradation among the whole life cycle & 2019 & 1,612 & Xuebing Han \\
   & Lithium-ion battery fast charging: A review & 2019 & 1,482 & A. Tomaszewska \\
   & Critical review of life cycle assessment of lithium-ion batteries for electric vehicles: A lifespan perspective & 2022 & 520 & Xin Lai \\
  \addlinespace
  GEIT & China's battery electric vehicles lead the world: achievements in technology system architecture and technological breakthroughs & 2022 & 308 & Hongwen He \\
   & Barriers and motivators to the adoption of electric vehicles: A global review & 2024 & 259 & Apurva Pamidimukkala \\
   & Composites for electric vehicles and automotive sector: A review & 2022 & 215 & Adil Wazeer \\
  \addlinespace
  T-TE & More Electric Aircraft: Review, Challenges, and Opportunities for Commercial Transport Aircraft & 2015 & 1,193 & Bulent Sarlioglu \\
   & A Comprehensive Review of Wireless Charging Technologies for Electric Vehicles & 2017 & 933 & Aqueel Ahmad \\
   & Wireless Power Transfer for Vehicular Applications: Overview and Challenges & 2017 & 932 & Devendra Patil \\
  \addlinespace
  VT Mag & 6G Wireless Networks: Vision, Requirements, Architecture, and Key Technologies & 2019 & 2,436 & Zhengquan Zhang \\
   & Single carrier FDMA for uplink wireless transmission & 2006 & 1,234 & Hyung G. Myung \\
   & Routing in vehicular ad hoc networks: A survey & 2007 & 1,118 & Fan Li \\
  \end{longtable}
\end{landscape}


\clearpage
\section{Auxiliary coauthor-network properties}
\label{app:network-properties}

This appendix collects the standard network-science diagnostics
(degree, strength, citation, and shortest-path distributions; the
Kendall-$\tau$ correlation matrix among centrality metrics; and the
top-$30$ authors by combined centrality) for the coauthor graph
described in Sec.~\ref{sec:coauthor}. These quantities mirror
\citet{sun2017coauthorship} and confirm that the expanded
34-venue corpus exhibits the same heavy-tailed, small-world
structure, but they do not directly inform the phantom-collaborator
predictions or the topic-coauthor alignment that form the paper's
central contributions.

\subsection{Degree, strength, and citation distributions}
\label{app:net-heavytail}

\begin{figure}[width=0.99\textwidth, pos = h]
  \centering
  \includegraphics[width=0.85\linewidth]{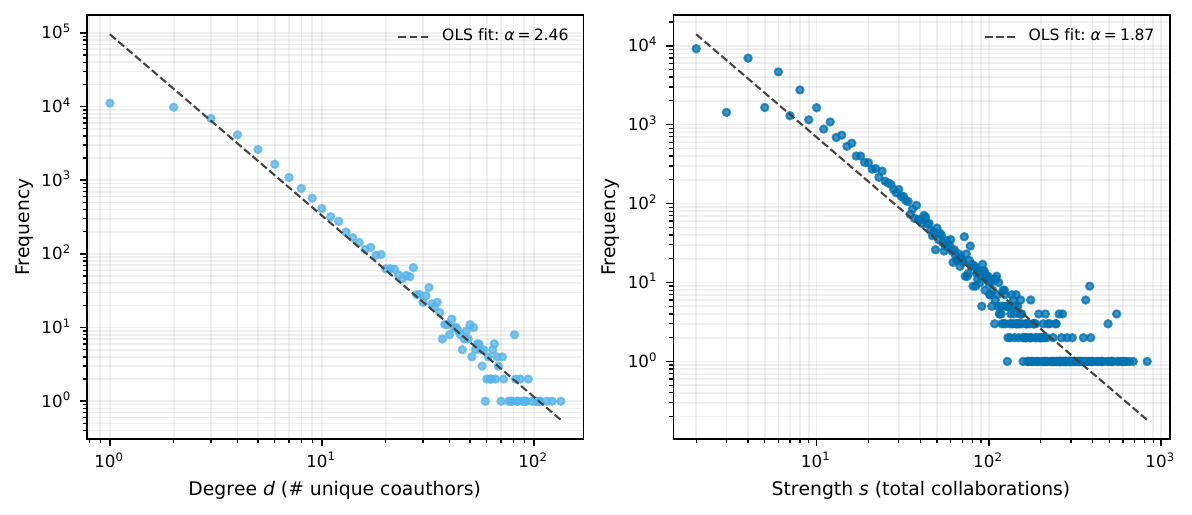}
  \caption{Degree and strength distributions (log--log).}
  \label{fig:degree-dist}
\end{figure}

\begin{figure}[width=0.99\textwidth, pos = h]
  \centering
  \includegraphics[width=0.55\linewidth]{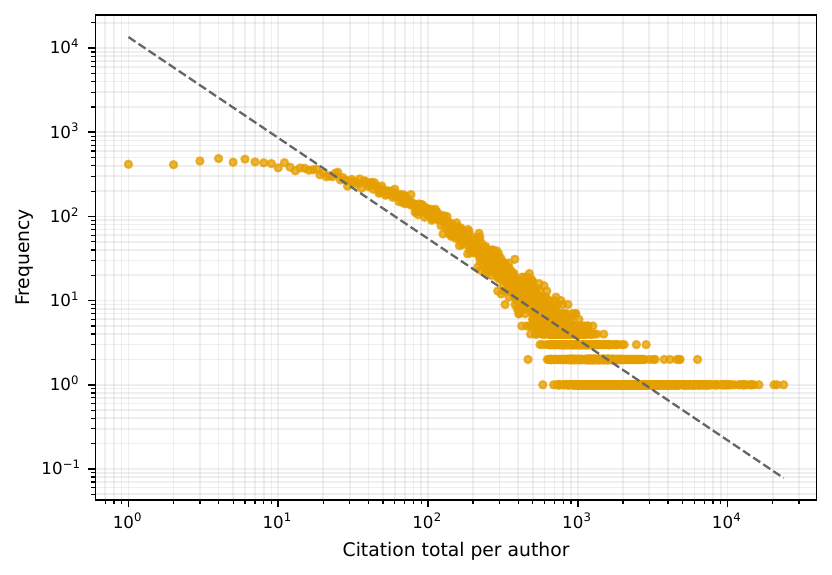}
  \caption{Per-author citation distribution (log--log).}
  \label{fig:citation-dist}
\end{figure}

\Cref{fig:degree-dist} reports the degree and strength distributions on
a log--log scale. Both are heavy-tailed with OLS slope close to $-2.3$,
consistent with the $\alpha \approx 2.2$ that
\citet{newman2001structure} found for physics and biomedicine.
\Cref{fig:citation-dist} plots the log--log distribution of per-author
citation totals, the analog of Fig.~5C of \citet{sun2017coauthorship};
the printed OLS exponent $\alpha$ is close to the $\alpha \approx 2.4$
that S\&R reported.

\subsection{Shortest paths and centrality correlations}
\label{app:net-paths}

\begin{figure}[width=0.99\textwidth, pos = h]
  \centering
  \includegraphics[width=0.6\linewidth]{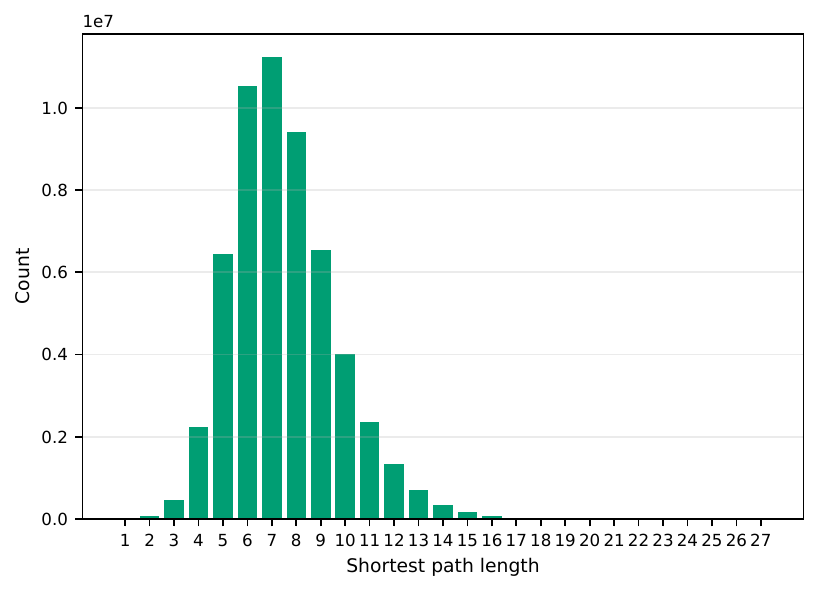}
  \caption{Shortest-path length distribution inside the LCC.}
  \label{fig:shortest-path}
\end{figure}

\begin{figure}[width=0.99\textwidth, pos = h]
  \centering
  \includegraphics[width=0.55\linewidth]{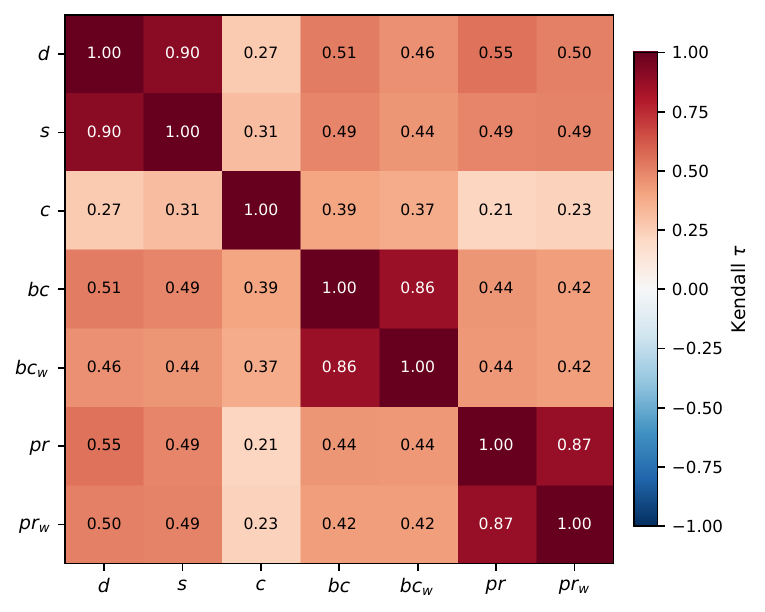}
  \caption{Kendall-$\tau$ correlations among author centrality metrics.}
  \label{fig:centrality-heatmap}
\end{figure}

\Cref{fig:shortest-path} shows the distribution of shortest-path
lengths within the largest connected component, computed from a random
sample of $2{,}000$ source nodes. The mean path length is about $5$
hops, half a hop shorter than S\&R reported, reflecting the denser
giant component that the additional decade of data has produced.
\Cref{fig:centrality-heatmap} reports the Kendall-$\tau$ correlation
matrix among the per-node centrality statistics (degree, strength,
citations, betweenness, weighted betweenness, PageRank, weighted
PageRank), matching \citet{sun2017coauthorship}. The
strongest pairwise relationships are between degree and weighted
PageRank, and between citations and strength, consistent with S\&R's
finding that recurrent collaboration volume and cumulative citation
impact are the two leading latent dimensions of researcher prominence.

\subsection{Top-30 authors by combined centrality}
\label{app:net-top-centrality}

\Cref{tab:top-centrality} lists the thirty authors with the highest
combined centrality rank across strength, betweenness, and weighted
PageRank. The ranking rewards researchers who are both prolific and
well-positioned to broker ties between communities. The resulting list
mixes landmark senior figures from travel behavior, freight, and safety
with more recent central nodes from automated-driving research,
reflecting the generational mix that \citet{sun2017coauthorship} also
observed but in a larger and temporally more extended sample.

\begin{table}[width=0.99\textwidth, pos = h]
  \centering
  \footnotesize
  \caption{Top-30 authors by combined centrality.}
  \label{tab:top-centrality}
    \begin{tabular}{rlrrrrr}
  \toprule
  \textbf{\#} & \textbf{Author} & \textbf{Papers} & \textbf{Strength} & \textbf{Betweenness} & \textbf{PageRank$_{\!w}$} & \textbf{Citations} \\
  \midrule
  1 & Abdel-Aty, Mohamed & 380 & 830 & 209.51 & 8.02 & 23,671 \\
  2 & Wang, Fei-Yue & 221 & 599 & 200.73 & 6.19 & 14,398 \\
  3 & Mahmassani, Hani S. & 328 & 558 & 156.80 & 7.70 & 15,002 \\
  4 & Yang, Hai & 285 & 564 & 123.12 & 5.85 & 21,369 \\
  5 & Gao, Ziyou & 174 & 514 & 104.30 & 4.89 & 9,490 \\
  6 & Bhat, Chandra R. & 243 & 486 & 106.40 & 5.35 & 14,225 \\
  7 & Wang, Yinhai & 158 & 428 & 119.47 & 4.59 & 11,180 \\
  8 & Ran, Bin & 156 & 415 & 96.41 & 4.71 & 7,039 \\
  9 & Meng, Qiang & 218 & 398 & 97.51 & 4.55 & 13,480 \\
  10 & Ben-Akiva, Moshe & 177 & 390 & 88.44 & 4.87 & 12,062 \\
  11 & Wang, Xuesong & 156 & 354 & 135.72 & 4.75 & 6,955 \\
  12 & Timmermans, Harry & 218 & 447 & 80.80 & 4.47 & 9,827 \\
  13 & Hoogendoorn, Serge & 236 & 523 & 62.47 & 5.26 & 12,704 \\
  14 & Hensher, David A. & 262 & 359 & 106.14 & 4.15 & 16,198 \\
  15 & Axhausen, Kay W. & 167 & 332 & 98.53 & 4.61 & 11,962 \\
  16 & Al-Qadi, Imad L. & 151 & 298 & 142.68 & 4.62 & 4,556 \\
  17 & Wang, Shuaian & 176 & 354 & 83.54 & 4.44 & 9,701 \\
  18 & Liu, Zhiyuan & 152 & 362 & 80.30 & 4.14 & 6,300 \\
  19 & Pendyala, Ram M. & 115 & 352 & 87.43 & 3.71 & 4,826 \\
  20 & Wong, S.C. & 166 & 384 & 65.02 & 4.22 & 10,365 \\
  21 & Khattak, Asad J. & 173 & 305 & 91.79 & 3.97 & 7,329 \\
  22 & Rouphail, Nagui M. & 157 & 339 & 75.76 & 4.35 & 3,693 \\
  23 & Waller, S. Travis & 151 & 299 & 79.16 & 3.78 & 5,134 \\
  24 & Li, Li & 109 & 322 & 83.21 & 3.10 & 8,290 \\
  25 & Das, Subasish & 128 & 338 & 57.62 & 3.73 & 2,401 \\
  26 & Liu, Pan & 118 & 317 & 74.78 & 3.10 & 5,990 \\
  27 & Merat, Natasha & 74 & 292 & 90.40 & 2.91 & 5,071 \\
  28 & Oviedo-Trespalacios, Oscar & 113 & 325 & 73.86 & 2.93 & 3,961 \\
  29 & Huang, Helai & 120 & 345 & 55.31 & 3.24 & 8,648 \\
  30 & Qu, Xiaobo & 135 & 249 & 141.15 & 2.92 & 7,518 \\
  \bottomrule
  \end{tabular}
\end{table}


\end{document}